\documentclass[fleqn,usenatbib]{mnras}

\usepackage{newtxtext,newtxmath}
\usepackage{deluxetable}
\usepackage{natbib,amsmath,multirow,tabularx}

\usepackage[T1]{fontenc}

\DeclareRobustCommand{\VAN}[3]{#2}
\let\VANthebibliography\thebibliography
\def\thebibliography{\DeclareRobustCommand{\VAN}[3]{##3}\VANthebibliography}

\usepackage{graphicx}	
\usepackage{amsmath}	
\usepackage{soul}

\title[Occulted Flares]{Identifying Flare Locations Through Exoplanet Transit Occultations}

\author[Armitage et al.]{%
        Tayt Armitage$^{1,2}$$^{\href{https://orcid.org/0000-0003-3190-8890}{\includegraphics[scale=0.5]{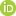}}}$,
        David V. Martin$^{3,2}$$^{\href{https://orcid.org/0000-0002-7595-6360}{\includegraphics[scale=0.5]{orcid.jpg}}}$,
        Romy Rodr\'iguez Mart\'inez$^{4,2}$$^{\href{https://orcid.org/0000-0003-1445-9923}{\includegraphics[scale=0.5]{orcid.jpg}}}$
\\
$^{1}$Department of Astronomy, University of Wisconsin-Madison, 475 N Charter St, Madison, WI 53706, United States of America\\
$^{2}$Department of Astronomy, The Ohio State University, 4055 McPherson Laboratory, Columbus, OH 43210, United States of America\\
$^{3}$Department of Physics \& Astronomy, Tufts University, 574 Boston Avenue, Medford, MA 02155, United States of America\\
$^{4}$Center for Astrophysics | Harvard \& Smithsonian, 60 Garden St, Cambridge, MA 02138, United States of America\\
\vspace{-0.3cm}
}

\vspace{-0.7cm}

\pubyear{2025}

\begin{document}
\label{firstpage}
\pagerange{\pageref{firstpage}--\pageref{lastpage}}
\maketitle

\begin{abstract}
M dwarfs are the most common stars in the galaxy, with long lifespans, a high occurrence rate of rocky planets, and close-in habitable zones. However, high stellar activity in the form of frequent flaring and any associated coronal mass ejections may drive atmospheric escape with the bombardment of radiation and high energy particles, drastically impacting the habitability of these systems. The stellar latitude where flares and coronal mass ejections occur determines the space weather that exoplanets are subject to, with high energy particle events associated with equatorial flares producing significant atmospheric erosion. However, the flaring latitudes for M dwarfs remain largely unconstrained. To aid in the effort to locate these flaring regions we explore the applicability of flare occultations using optical photometry to identify the latitudes of flares. As a planet transits in front of an ongoing flare the timing and geometry of the transit can be used to constrain the latitude and longitude of the flare. We predict the probability of detecting an occultation for known transiting planets and eclipsing binaries. From this, we estimate 3--22 detectable occultations exist within the TESS primary mission photometry, with the majority occurring in eclipsing binary observations. To demonstrate this technique, we analyze a candidate flare occultation event for the eclipsing binary CM Draconis.

\end{abstract}


\section{Introduction}\label{sec:introduction}

A priority area of modern astronomy is the characterization of exoplanets and their atmospheres \citep{2021pdaa.book.....N}. In this search, M dwarfs have become very popular targets. This is in part due to being the most common type of star in the galaxy, but they also have a high occurrence rate of rocky planets \citep{Howard:2012,Dressing2015,Mulders:2015,Hardegree-Ullman2019}. In addition to this, many of the Habitable-Zone (HZ) planets, which are defined as orbiting in the region where liquid water may exist on the planet's surface \citep{Kasting1993,Shields2016}, have been found around M dwarfs (e.g., \citealt{Crossfield:2015,Anglada-Escude:2016,Gillon:2017, VanEylen:2021,Gilbert:2020}). 

However, the high flare activity of M dwarfs  (e.g., \citealt{Kowalski:2009,Hawley:2014, Davenport2014, Schmidt:2019, gunther2020, RodriguezMartinez:2020,Feinsten:2020}) calls into question the habitability of the large population of temperate terrestrial planets in these systems. A flare is defined as a sudden brightening of a star in which a burst of radiation is released \citep{10.1093/mnrasl/slab055}. This radiation then poses a threat to the habitability of planets in these systems from the frequency of flaring on M dwarfs. This is made worse as these flares are typically more energetic than those occurring on our own Sun. Flares can have drastic impacts on the potential habitability of exoplanets around M dwarfs \citep{Lammer:2007,Segura:2010,Tilley2019, Chen:2021}. One of the core components of habitability is the ability to retain a sufficient atmosphere alongside liquid water. In the presence of high energy flares, constant bombardment can lead to significant atmospheric erosion and high rates of water loss \citep{doAmaral2022,2020AJ....160..237F}. Flares also directly alter the chemical composition of an exoplanet's atmosphere. Such reactions include the production of ozone, which would mitigate the effect of future flares, or the abiotic production of chemicals typically used as bio-signatures like N$_{2}$O \citep{Ridgeway2023}. The release of high energy photons from a flare can be considered isotropic, and will therefore reach any planet on the same side of the star during the flare event. However, the flare energy that hits the planet will be attenuated by a self-shadowing as a function of the stellar latitude \citep{Ilin2021}, such that a polar flare will impart $\approx 20\%$ the energy of an equatorial flare.

In addition to flare radiation, planets are also bombarded with high energy particles from coronal mass ejections (CMEs). A coronal mass ejection is an expulsion of highly energized and magnetized plasma from the surface of the star. This expulsion is directional, as opposed to the omnidirectional release of radiation from flares. It is believed that CMEs and flaring regions are correlated \citep{doi:10.1016/j.nrjag.2012.12.014}. This relationship is not fully understood even on the Sun, as not every flare comes with a CME and vice versa. Still, we must consider them nonetheless, particularly since the latitude of CME emission has a strong effect on habitability. The severity of the radiation is increased if the exoplanets have little to no magnetic fields to protect themselves (e.g., \citealt{Driscoll:2015}). In this situation, it can lead to extreme or total atmospheric loss \citep{Tilley2019}.
However, if the atmosphere is retained, the high energy particles can aid in the production of molecules crucial for hosting life \citep{namekata2022hunting}.

In order to properly gauge the impact of flares and CMEs, it is imperative that we know their locations on the star. On our own Sun, flares occur mostly within $30^{\circ}$ of the equator \citep{Balasubramaniam:1994,Chen:2011}. Flares emitted primarily from equatorial latitudes mean that planets aligned with the stellar spin axis would be regularly bombarded with energetic particles and radiation. The assumption that the planet's orbit is aligned with the stellar spin applies to the Solar System and most exoplanets, although some exoplanets orbit with high inclinations\citep{Winn:2015}.
However, given that late M dwarfs are fully convective and hence have a very different stellar structure, we cannot necessarily translate solar behavior to M dwarfs. There is a chance for other flare distributions, such as polar flaring. If polar flaring were to occur, most of the radiation and energized particles released from CMEs would be directed away from exoplanets, minimizing the harmful impacts on their atmospheres. Thus, it is crucial to constrain the location of flaring regions on M dwarfs. 

Current efforts to localize flaring regions on M dwarfs are quite limited. A promising avenue is the photometry of rapidly rotating M dwarfs, where the rotation rate is shorter than the flare duration. This causes a modulation of the flare signal as it passes in and out of the line of sight of the observer. Modeling this effect reveals the flare's latitude. \citet{Ilin2021} applied this to six flares on four M dwarfs and calculated latitudes between 50 and 80$^{\circ}$. \citet{Ilin2022} reviewed the known (at the time) flare localization methods: indirectly, with light curve inversion used in Doppler imaging and asteroseismology, and directly, with the aforementioned rotation-modulation. In addition, \citet{ILIN:2023} explores the possibility of using the waiting time between flares on active stars to infer the latitudes of active regions. This is done by defining the waiting time as a function of phase and comparing it across a sample of 200 stars.

Work to identify flares using occultations has also been done with X-ray flares. A flare occultation event occurs when a transiting body \footnote{Transiting planet or eclipsing binary, but we will typically use the word transit.} passes in front of an ongoing flare blocking out some of the light.\citet{2006A&A...445..673S,2007A&A...466..309S} analyze used flare occultations at soft X-ray wavelengths to constrain the latitudes of active regions in multiple eclipsing binary systems. The authors then detailed their techniques to model the geometry of the systems in \citet{2007A&A...466..309S}, and the reported the latitude ranges for each flare. Of the flares analyzed, one observed in 
\citet{1999Natur.401...44S} was constrained to polar latitudes. The rest were limited to lower regions of their host stars. 

In this paper, we move beyond the X-ray regime and explore the applicability of localizing flares through the identification of flare occultation events in optical datasets. This does provide more difficulty as white light flares display complex substructures as detailed in \citet{2022ApJ...926..204H}, which are absent from the soft X-ray flare population. However, there is a benefit to using this population as over a  million flares have been detected across optical sky surveys \citep{Feinstein_2022,2023A&A...669A..15Y}. To explore this method we focus on systems observed using the Transiting Exoplanet Survey Satellite (TESS; \citet{Ricker:2015}. We propose that these events can be used to localize the region the flare occurred using the known planetary information. This is done by knowing the inclination of the star and the impact parameter of the system. The impact parameter is what describes the perpendicular distance of the exoplanet relative to the center of the star it orbits. Using these parameters, we can determine the latitude of the transiting planet relative to the host star. With an occultation event, we can then infer the location of the flare as it would have to be restricted to the latitudes of the star that are blocked by the transiting exoplanet. With a significant amount of occultations identified a general distribution of the flare activity for M dwarfs could be developed. We are directly constraining flare latitudes, whereas any constraints to the CME's latitude are done indirectly through the presumed relationship between the two events.

Our paper is organized as follows. First, we develop the theory behind our observational technique by modeling what an occultation would look like (Sect.~\ref{sec:modelling}). Next, we predict the likelihood of observing such an event (Sect.~\ref{sec:probability}). We determine the relationship certain parameters may have on this probability and calculate the probability for four well-known exoplanet and eclipsing binary systems (Sect.~\ref{sec:parameter_effects}). Following this, we put forward how this technique may be used with observations (Sect.~\ref{sec:methodology}), including a candidate occulted flare in CM Draconis (Sect.~\ref{subsec:CMDraconis}). Then we assess the available dataset and how many occultations can be expected to be found with no new observations (Sect.~\ref{sec:assumptions}). Finally, we discuss the implications of our results and scenarios that may impact this technique's efficacy (Sect.~\ref{sec:discussion}) before concluding (Sect.~\ref{sec:conclusion}).

\section{Modeling Occultation Events}\label{sec:modelling}

In order to localize the flaring regions on M dwarfs, we have developed a simulation that models the geometry of a planet transiting across a star, with an occultation occurring as the planet passes in front of a flare. It then checks if a flare occultation occurs and uses the geometric information to simulate the light curve that would be produced from the occultation. These simulations will serve as the baseline for identifying genuine occultations. 

\subsection{Transit Simulation}

To predict the morphology, or shape, of a flare occultation, our simulation first models the geometry of the transit/eclipse. In order to model the transit, our simulation requires a few inputs: impact parameter, period or transit duration, the radius of the host star, the radius of the planet or secondary star, and the cadence of observation. Utilizing the impact parameter, we are then able to determine the latitude of the planet/eclipsing binary as it sweeps across. This range of latitudes will then encompass the overall range at which we will be able to constrain flaring regions. Our simulation then models the transit on a 2D plane, with the star and planet represented as circles. The speed at which the transit occurs is either determined solely by the prior transit duration input or determined using the period of the system. The simulation then utilizes the cadence data, given in seconds, to precisely map out the position of the transiting/eclipsing body at each point in observation. It is important to model it this way rather than a continuous sweep, as cadence sensitivity also plays an important role in our ability to be able to observe an occultation event. 

\subsection{Simulating Occultation Geometry}

 While the transit is being simulated, our model is simultaneously simulating flares occurring on the host star. The amount of flares occurring during a given modeled transit is determined by an input white light flare frequency for the host star.
 Our model then randomly generates the values of the flares' amplitude, full-width half maximum of the flare duration, peak time, and radius of the flaring region. In order to do this, we generate a seeded value from a normal distribution. This seed then sets what the other parameters of the flare will be, with the most common as low-energy flares but still factoring in rare, very high-energy flares. The first parameter determined is the amplitude of the flare. The flare's amplitude is represented in the flux of flare relative to the normalized flux of the star, with our model considering 3\% to 20\% increases in flux from a single flare event. The full width half maximum values and the radii of the flares are then randomly generated in different thresholds based on the value of the amplitude. The full-width half max determines the duration of the flare, with low amplitude flares having durations of 10-45 minutes, while high amplitude flares may last from 2-4 hours in our simulation. The simulation incorporates a predicted radius of the flare, ranging from an Earth radius to a Jupiter radius in the case of a super-flare. The underlying assumption is that higher energy flares will a larger area. The flares are modeled as a circular region on the host star's surface. The position of the flare is also generated at random.
 When generating the positions of the flares, we operate on the assumption that flares occur uniformly across M dwarf stars. We assign flare positions across the spherical surface of the star and then represent them on the same 2D grid in which we model the transit.
 
 Once the parameters for the flares have been determined, the simulation then keeps track of the area of the flare that is visible at all points of observation. While modeling the transit, the flare is present for the entire duration. However, this is exclusively for the geometric calculation, as when checking for an occultation, we account for the peak time and duration of the flare. The model then keeps track of the position of the transiting body relative to the flare at all points of observation. If the distance between the flare's center and the planet's center is small enough such that the planet covers part of the flare, the occulted area is then calculated using the following equation adapted from \citet{Weisstein_2003}:
 \begin{equation}
 \begin{split}
     a =  R_{\rm f}  \cos^{-1}\Biggl(\frac{d^2 + R_{\rm f}^2 - R_{\rm p}^2}{2d R_{f}}\Biggl)  ~+~ R_{\rm p}^{2} \cos^{-1}\Biggl(\frac{d^2 + R_{\rm p}^2 - R_{\rm f}^2}{2  d  R_{\rm p}}\Biggl)  ~-~ \\
     \frac{1}{2} \Bigl((-d + R_{\rm f} + R_{\rm p}) (d + R_{\rm f} - R_{\rm p}) (d - R_{\rm f} + R_{\rm p})  (d + R_{\rm f} + R_{\rm p})\Bigl)^{1/2}.
 \end{split}
 \end{equation}
 Where $a$ represents the total area occulted, $R_{\rm f}$ and $R_{\rm p}$ represent the radius of the flare and the planet, respectively, and $d$ represents the distance between the center of the flare and the planet.

Once the occulted area has been calculated, our model then finds the fraction of the flare that would still be visible and records this. From these measurements, we are able to graph the geometry of the occultation while also displaying the position of the event on the star.
 
\subsection{Determining an Occultation Event}
The area calculations defined above are critical to determining whether or not an occultation has occurred. Within our simulations, the flares are considered as always present, meaning that there should initially be 100\% of their area showing. We take on the notion that for an occultation to occur, there must be a 10\% dip in the flux produced by the flare. While we could record any decline in flux as an occultation, we set this limit to reflect the occultations that we would be able to identify in real data, i.e. we only consider detectable occultations. Therefore, flares that have had 10 \% of their area or more occulted at any point on the transit are flagged for further analysis within the simulation as possible occultations, while the rest are no longer analyzed. 

Then we take into consideration the peak time and duration of the flare. To ensure that there are no false occultations, being those that occur well before the peak time, we set a threshold for when they can occur, taking both the peak time and flare duration into account. For all flares, we first ensure that the occultation occurs within the eclipse time, so we only consider the decreases in the total area shown for data points that occur at or after the peak time of the generated flare. We break up the optimal window of an occultation occurring by the full-width half maximum of the flare duration, setting the threshold as occurring at a time at most 4 times this value after the peak time of the flare. After this threshold, the flux from the flare has decreased enough that it would not be detectable in real-world observations. 

In addition to these checks, there is also a means to root out any flares that had been completely occulted for their duration, as the initial criteria of such an event would satisfy the above conditions. The solution was to ensure that the simulation checks that in the event of an occultation, there is at least one point where the flare would be visible while it is still ongoing. As such, we would have a chance to observe some change in flux rather than a constant zero line indicating the flare was completely occulted for the entirety of the flare's duration and would thus not appear in actual photometry.

 
\subsection{Visualizing the Light Curves}

When our simulations identify that a flare occultation event has occurred, we are then able to create a modeled light curve to simulate the morphology of the light curve if we were to observe it. To generate the initial light curve of the flare, we incorporate flare modeling morphology developed in \citet{Mendoza2022}, which is an improvement on the classic equation from \citet{Davenport2014}. The flare morphology as a function of scaled time $f(t^{*})$ is given by

\begin{equation}
\label{eq:flare_profile}
f(t^{*}) = \frac{\sqrt{\pi}AC}{2}\left[F_1h(t^{*},B,C,D_1) + F_2(t^{*},B,C,D_2) \right],
\end{equation}
where
\begin{align}
\begin{split}
h(t^{*},B,C,D) &= \exp\left[{-Dt^{*}+\left(\frac{B}{C} + \frac{DC}{2}\right)}\right]^2\times {\rm erfc}\left(\frac{B-t^{*}}{C} + \frac{DC}{2}\right) 
\end{split}
\end{align}
and
\begin{equation}
{\rm erfc} = \frac{2}{\sqrt{\pi}}\int e^{-s^2}ds
\end{equation}
is the error function for which we use the Scipy package \textsc{scipy.special.erf}. The fixed variables in Eq.~\ref{eq:flare_profile} are the flare profile coefficients from \citet{Mendoza2022}:

\begin{align}
A &= 0.9688\\
B &= -0.2513\\
C &= 0.2268\\
D_1 &= 0.1555\\
D_2 &= 1.2151\\
F_1 &= 0.1270 \\ 
F_2 &= 1 - F_1
\end{align}

Our input parameters for the flare are $t_{\rm peak, flare}$ (time at which the flare peaks), ${\rm FWHM}_{\rm flare}$ (duration of the flare, expressed by a full-width half maximum) and $A_{\rm flare}$ (amplitude of the flare). We determine the flare flux  as a function of time by 

\begin{equation}
    {\rm flux}_{\rm flare}(t) = f(t^{*}) \times A_{\rm flare},
\end{equation}
where $t^{*}$ is the scaled time to incorporate our input ${\rm FWHM}_{\rm flare}$:
\begin{equation}
    t^{*} = \frac{t - t_{\rm peak, flare}}{{\rm FWHM}_{\rm flare}}.
\end{equation}

Then in order to incorporate the occultation into the light curve, we combine the calculated area that remains visible with the flux generated by the flare model. This then produces a visible decline in flux as the flare is occulted, which we have seen to produce many differing flare morphologies. 

\begin{figure*}
    \centering
    \includegraphics[scale = 0.5]{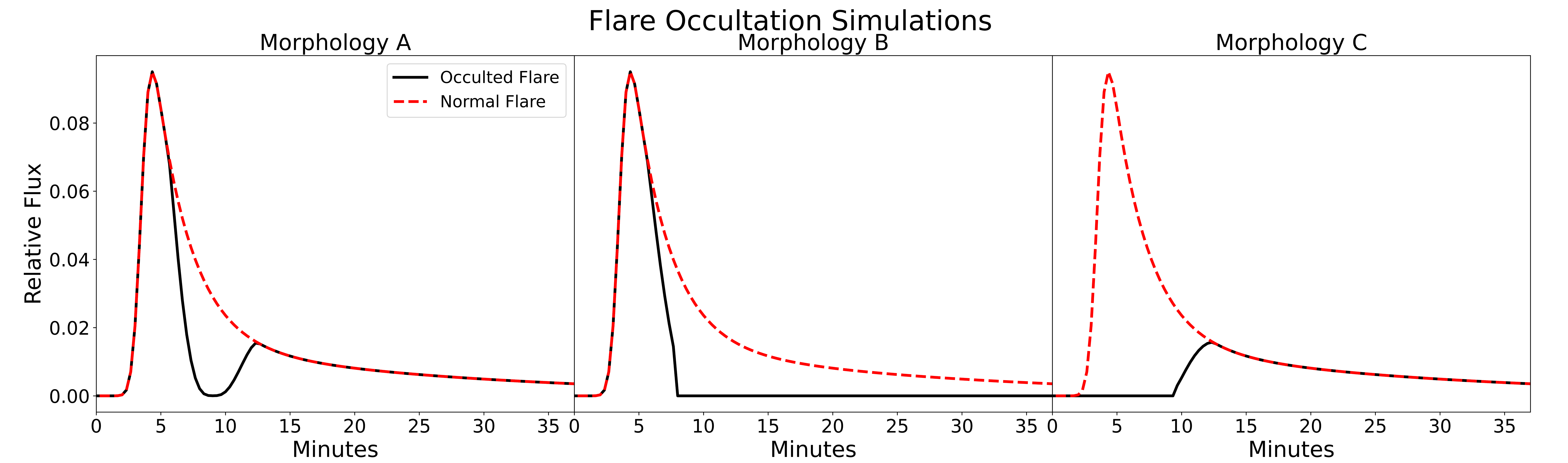}
    \caption{Three examples of possible flare occultation morphologies that have been simulated. In these events, the occultation occurs for part of the decay phase, the entire decay phase, or during the rising phase to produce unique light curves. In Sect.~\ref{subsec:caveats}, we discuss the caveats of how these different morphologies could be produced without occultations.}
    \label{fig:flare_examples}
\end{figure*}

We can see in Fig.~\ref{fig:flare_examples} that many morphologies can be produced from an occultation event. Fig.~\ref{fig:flare_examples}A occurs when the flare is occulted during the decay phase causing a dip in the flux. The occultation then ends while the decay phase is ongoing, resulting in an increase in flux. Such a signal would be very useful for identifying occultation events and would occur more frequently with smaller bodies or during longer-lasting flares that occur during shorter eclipses. Fig.~\ref{fig:flare_examples}B demonstrates a morphology that can occur when a flare is occulted during the decay phase, which is shorter than the occultation event. This results in a sharp rise in flux that is immediately cut off to zero as the flare is completely occulted. The third morphology pictured in Fig. ~\ref{fig:flare_examples}C is what can occur if the occultation event occurs during the rising phase of the flare. This morphology does not deviate greatly from the standard shape of a classical flare, making it quite difficult to differentiate between this type of event and less energetic flares with lower amplitudes. Therefore in observation, we would seek the first two morphologies pictured as the third type too closely resembles a low amplitude flare to be distinguishable as an occultation. 

\section{Calculating the Occultation Probability}\label{sec:probability}
In this section, we demonstrate how likely an occultation event is to occur for various systems. This is important as these events could be quite rare. Hence, it is imperative that we know which parameters influence this probability, especially where they would increase the likelihood of seeing an event. 

To accomplish this we use our occultation simulation, which has the ability to predict the probability of observing a flare occultation event for a single transit. This is done through the use of a Monte Carlo simulation, where we simulate a single transit/eclipse 50,000 times. The first check in this simulation is whether or not a flare had occurred during the transit, and we determine this probability as \\
\begin{equation}    
\rm P_{Flare} = {flare~ frequency \times transit~ duration}.
\end{equation}
As we simulate each transit, a seed is generated and checked against this probability to determine whether or not flares are generated in the given run. If no flares are generated it is automatically deemed a failed run, and the next simulation begins. If flares are generated then the transit is simulated, and the geometry is checked. The transit is deemed successful if at least one flare is occulted during the simulation. 
The number of successful and failed observations are stored, where at the end of the simulation, the probability of observing a flare in a single transit is found as 
\begin{equation}
\rm P_{occultation} = \frac{N_{success}}{N_{runs}}.
\end{equation}
\\

With this initial probability, we can then extend it to longer observing runs by finding the probability of finding at least one occultation event across many transits in longer observing runs. This probability is given by

\begin{equation}
\rm P_{obs} = 1 - (1-P_{occultation})^{N_{transit}}  .
\end{equation}
Using this probability, we can then predict the number of 
transit observations required to confidently find an occulted flare under our assumptions. 

On the Sun, flares occur mostly at latitudes between $\pm30^{\circ}$ and there is preliminary evidence that flares from rapidly-rotating M dwarfs occur closer to the poles \citep{Ilin2021}. In our models, we assume that flares occur following a spherically-uniform distribution on the surface of the star. This allows us to isolate the effects of different orbital parameters on the occultation probability.

\section{Impact of various parameters on observation probability}\label{sec:parameter_effects}

In this section, we explore how transit parameters impact the likelihood of observing an occultation event. By analyzing various individual parameters we can gauge the most ideal conditions for witnessing an occultation event. To demonstrate this we simulate transits using Earth-sized, Neptune-sized, and Jupiter-sized planets in a range of different conditions. These simulations will then aid in choosing targets and cadences for future follow-up observations.

\subsection{Probability vs Radius Ratio}
The first parameter that we explore is the effect that the radius ratio of the transiting body to the host star has on the probability of observing an occulted flare. The larger the radius, there is a greater area of the host star covered during transit, allowing for more opportunities of occultation.

\begin{figure}
    \includegraphics[width = 0.5\textwidth]{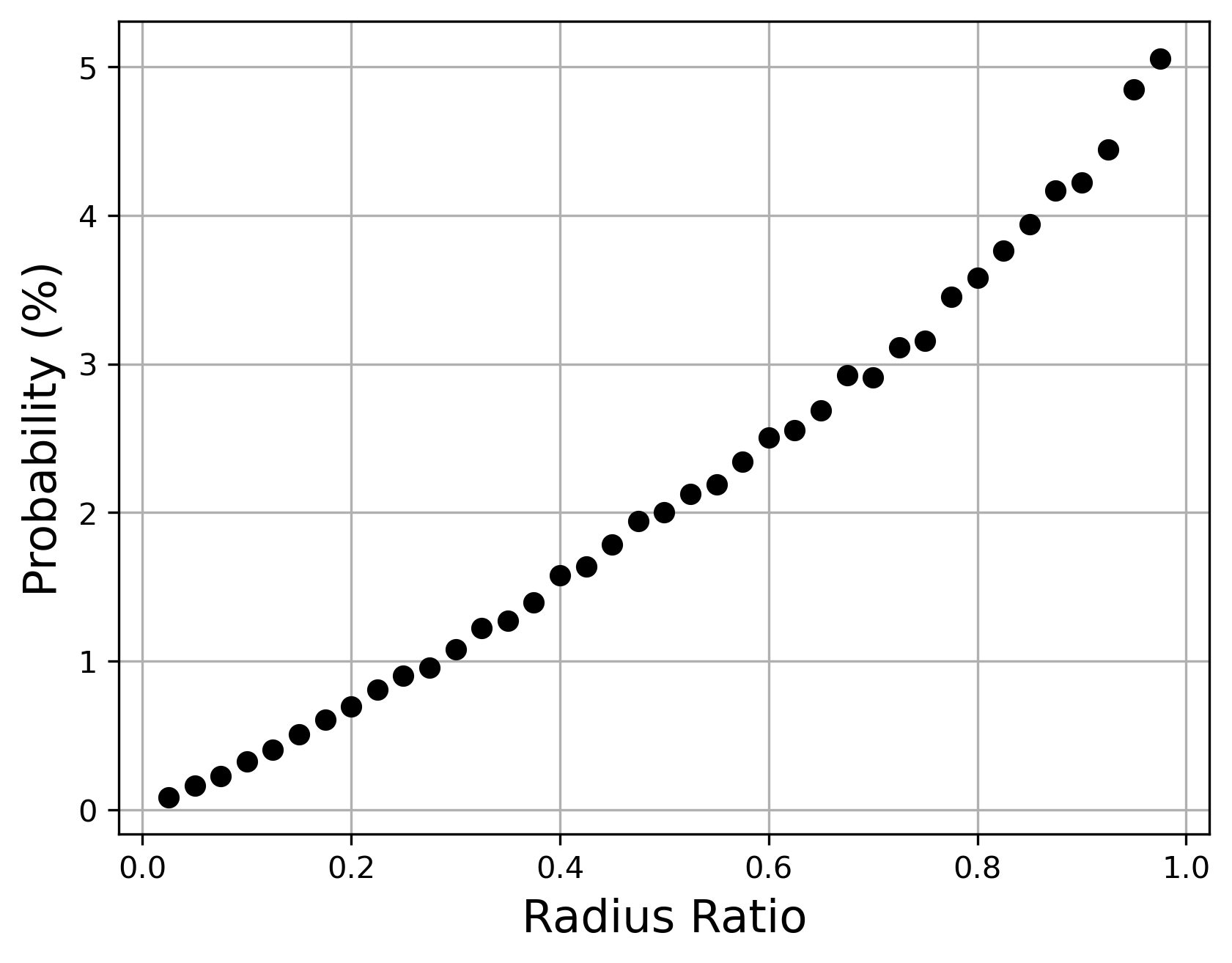}
    \caption{ Single transit flare occultation probability as a function of radius ratio. In these simulations, the only parameter that varied was the radius of the orbiting body, $R_{\rm p}$. Base simulations ran this transit 50,000 times assuming initial parameters of $R_\star = 0.25$ $R_\odot$ as a benchmark, $b = 0$, $P = 7$ days, and a flare rate of 3 flares per day.}
    \label{fig:probability_Radius_ratio}
\end{figure}

In Fig.~\ref{fig:probability_Radius_ratio}, we detail the effect of these different radius ratios.

We find that there is a direct correlation between the radius ratio and the probability of observing an occultation event. This indicates that larger planets or eclipsing binaries would be particularly advantageous in detecting occulted flares. However, there is a trade-off with precision with larger radius ratios, which is discussed in detail in section 6.

\subsection{Probability vs Impact Parameter}

\begin{figure}
    \centering
    \includegraphics[width = 0.5\textwidth]{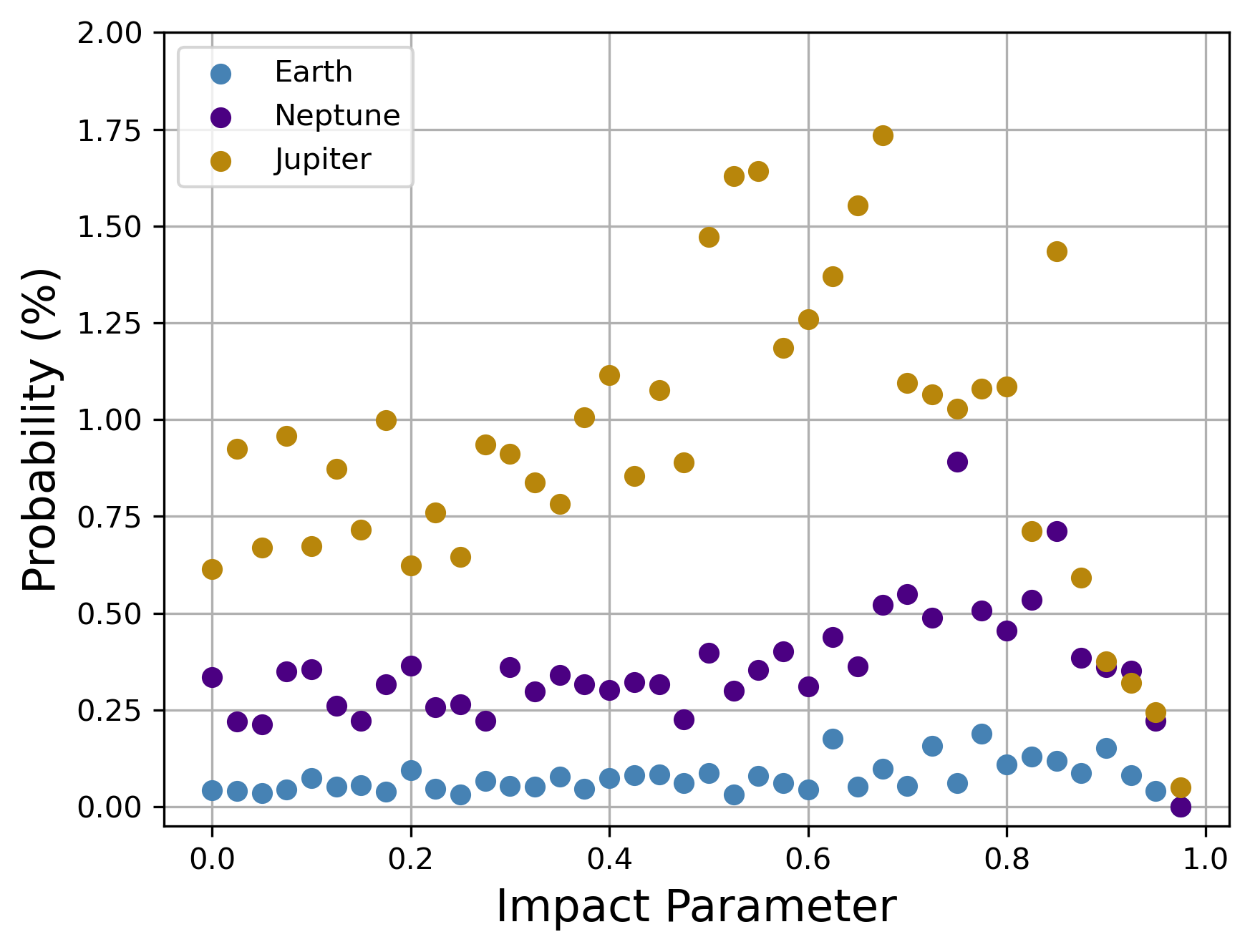}
    \caption{ Single transit flare occultation probability as a function of impact parameter, calculated for four different radius ratios. Again we use base parameters of $R_\star = 0.25R_\odot$, a 7-day period, and a flare rate of 3 flares per day, simulated across 50,000 transits. The graph also includes multiple radii including planets of $R = 1R_\oplus$, $R = 1R_{\rm Nep}$, and $R = 1R_{\rm Jup}$.}
    \label{fig:probability_b}
\end{figure}

The impact parameter dictates the latitudes that we are able to observe for flare occultations, it is imperative that we know if there are any observational biases based on different impact parameters. Pictured in Fig.~\ref{fig:probability_b}, we detail the effect of different impact parameters on the probability of detecting an occulted flare. 

In Fig.~\ref{fig:probability_b}, we can see interesting behaviors regarding the impact parameter on probability. For planets that have a lower radius ratio, the probability of occultation is not greatly affected by the impact parameter. There is a slight upward trend at higher values up until those of nearly 0.9, where the transit only grazes the star. At this point, the probability of occultation begins to decrease as less of the planet's area is covering the star, and thus there is less area that can occult flares. For larger planets 

the probability of occultation increases more dramatically as the impact parameter grows until large impact parameters where the probability begins to drop off. The higher probability of occultation is likely a selection bias as higher latitudes of the stars will appear to have higher concentrations of flares when projected onto 2D planes as the curvature of the star is more extreme in those locations. The drop off in probability seen following the upward trend is a result of the planet grazing the star, leaving less area swept over, decreasing the chance of a flare occultation event. This effect does not impact low radius ratio systems as the planet as greatly as they cover a smaller portion of the star that would be impacted by the 2D projection, so only a small upward trend in probability is seen, but the issue arising with grazing transits still applies.

\subsection{Probability vs Flare Rate}
The next parameter we explore is the effect flare rates may have on observations. M dwarf stars display have a wide range of flare activity, thus it is important to understand how significantly differing flare rates impact occultation probability.

\begin{figure}
    \includegraphics[width = 0.5\textwidth]{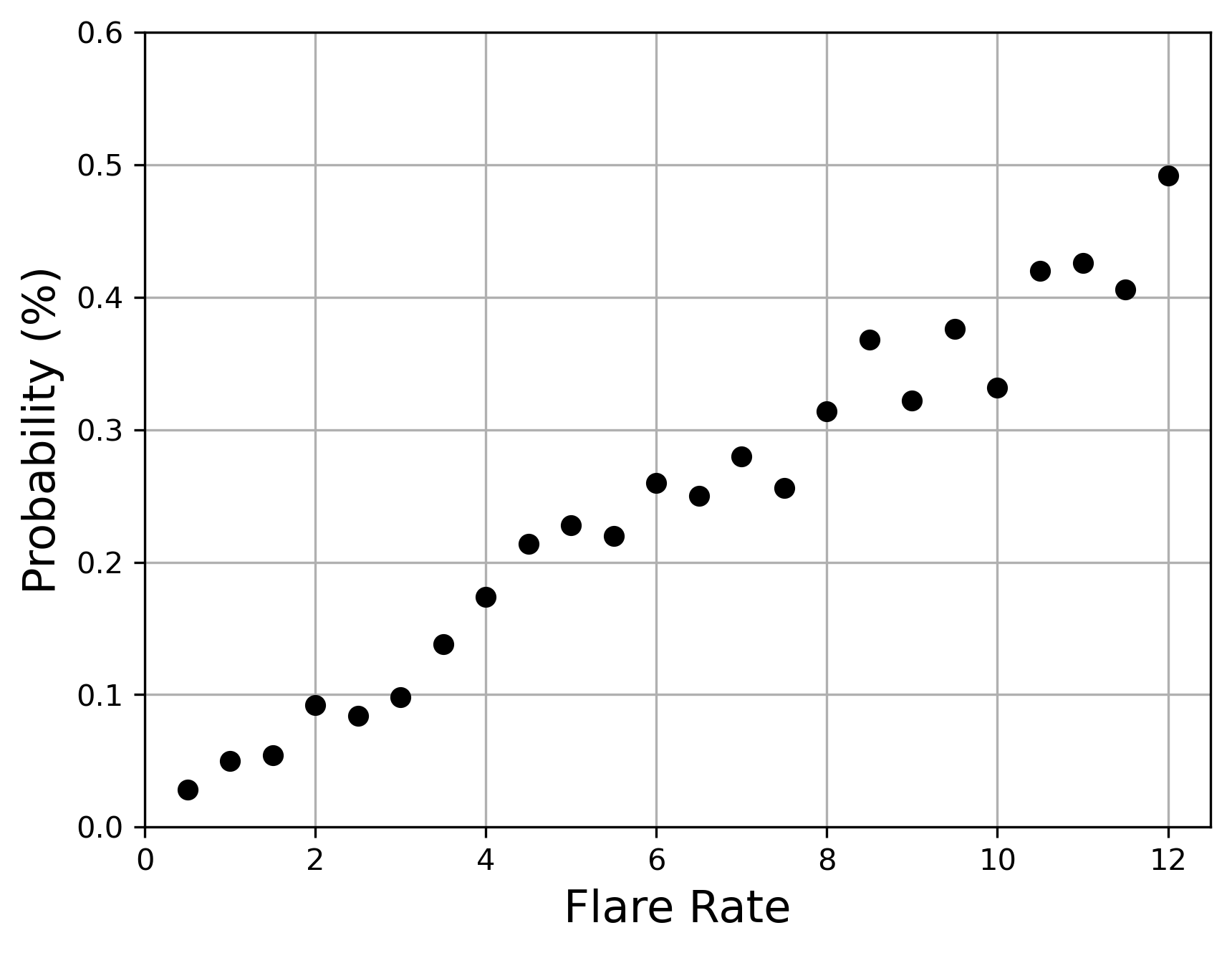}
    \caption{Flare occultation probability as a function of flare rate. Aside from the flare rate, the simulation parameters are fixed at $R_\star = 0.25R_\odot$, $R_{\rm p}=1R_\oplus$, $P = 7$ days, $b=0$. }
    \label{fig:probability_flare_rate}
\end{figure}

Viewing Fig.~\ref{fig:probability_flare_rate}, we can see that the likelihood of observing an occultation follows linearly with the flare rate of the host star. This follows as more flares equate to more chances of catching an occultation. Cases in Fig.~\ref{fig:probability_flare_rate} where the probability does not increase monotonically can be attributed to statistical noise in the Monte Carlo simulations. While a higher number of simulations can be done to eliminate this noise, we choose to remain at 50,000 as it keeps the run time at a manageable level while still producing trends.

\subsection{Probability vs Orbital Period}
 
We now look at the effect orbital period can have on the probability for occultation.  This is particularly important as the period of a system will influence the number of observations that are available to search for flare occultations.

\begin{figure}
    \includegraphics[width = 0.5\textwidth]{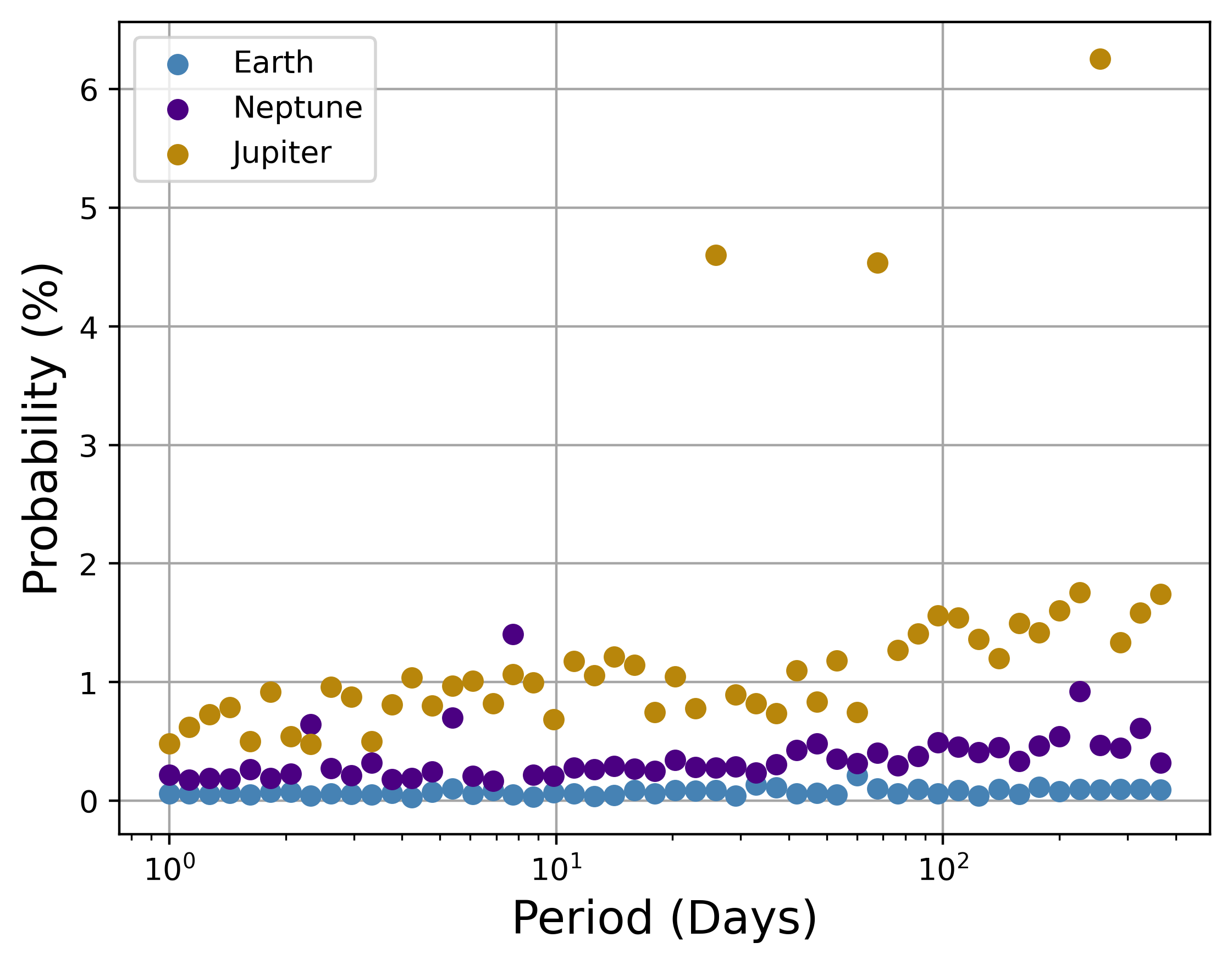}
    \caption{Single transit flare occultation probability as a function of the planet's orbital period. Again we use base parameters of $R_\star$ = 0.25R$_\odot$, a flare rate of 3 flares per day, $R_{\rm p} = 1R_\oplus, 1R_{\rm Neptune}, 1R_{\rm Jupiter}.$}
    \label{fig:period}
\end{figure}
\begin{figure}
    \includegraphics[width = 0.5\textwidth]{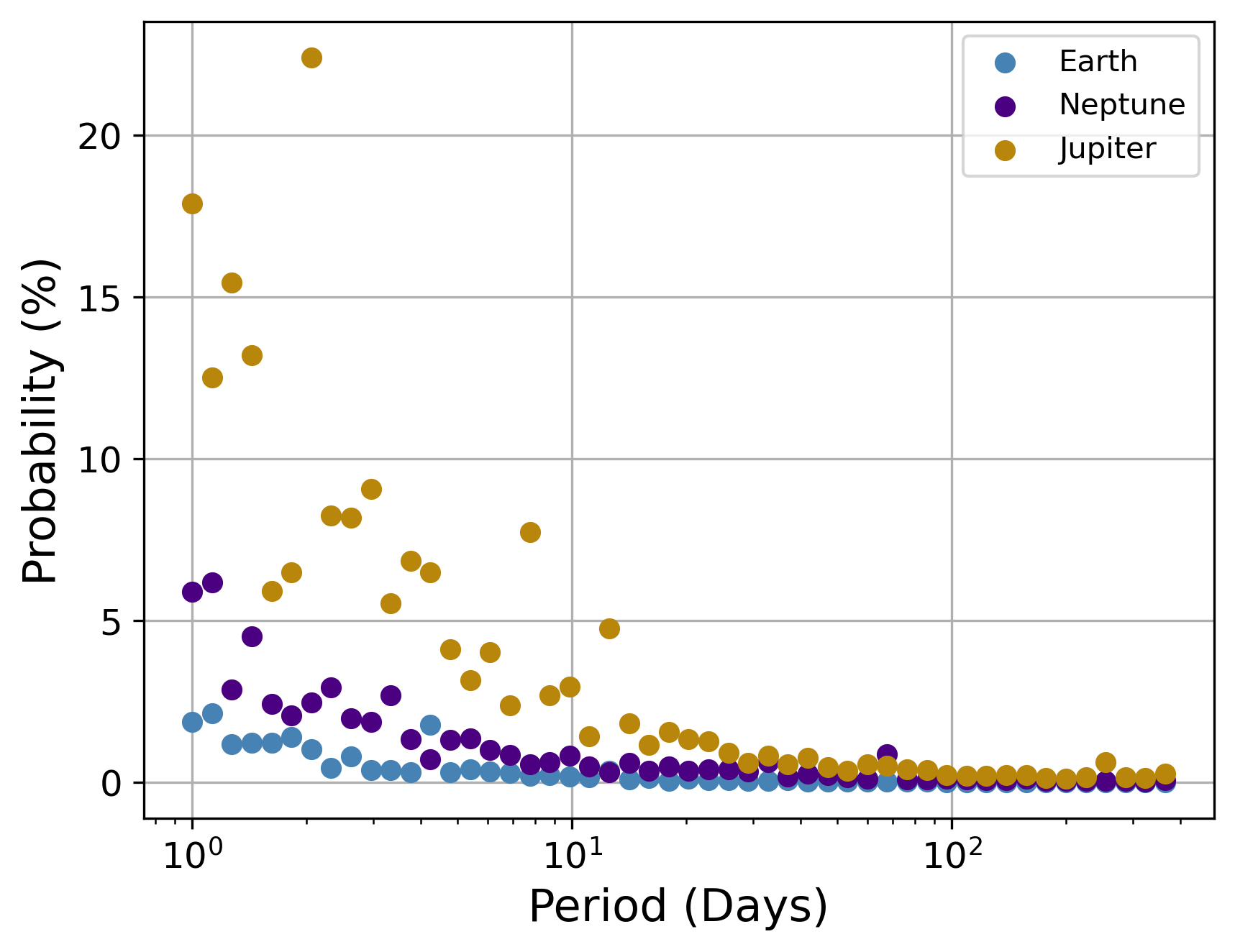}
    \caption{Flare occultation probability during an entire TESS sector as a function of the planet's orbital period. Again we use base parameters of $R_\star$ = 0.25R$_\odot$, a flare rate of 3 flares per day, $R_{\rm p} = 1R_\oplus, 1R_{\rm Neptune}, 1R_{\rm Jupiter}$. Even though the occultation probability for a given transit does not strongly depend on orbital period (Fig.~\ref{fig:period}), a shorter period planet has more transits and hence more chances for an occultation.}
    \label{fig:TESS_Sector}
\end{figure}

When analyzing the effect of orbital period on the probability of occultation we first considered the observation of only a single transit. For our sample size, we chose periods ranging from 1 day to 365 days, with 50 data points spaced evenly on a logarithmic scale. This was done using Earth-sized, Neptune-sized, and Jupiter-sized planets. We see in Fig.~\ref{fig:period} that there is very little effect on the single transit probability when considering differing periods for the smaller planets. For the Jupiter-sized planet, the effect becomes more noticeable where a larger period does yield a higher probability of detection for a single transit. This higher probability for a single transit is the result of an extended transit duration that would give way more opportunities for an occultation. 

With short-period planets, it seems unfavorable for a single transit, but there will be more opportunities for observations in a single TESS sector. Fig. ~\ref{fig:TESS_Sector} displays this change as we portray the individual transit probabilities scaled with the number of transits we would be able to witness in a single TESS sector. While the single transit probability for short-period planets is low, planets that had a >1\% probability of occultation in a single transit have detection probabilities upwards of 14\% across the entire observation window.
The probability of occultation quickly decreases as the periods grow larger until there is little to no change from the single transit probability as the period nears and then surpasses the duration of the TESS sector. 

\subsection{Probability vs Cadence}
Across the different TESS sectors, there were a number of observational cadences used, which could have an important effect on our ability to detect occultations. Therefore, in order to explore this, we ran simulations with cadences utilized by the TESS survey. Standard TESS observational cadences occur at 20 second, 120 second, 200 second, 10 minute, and 30 minute intervals. 
In addition to this, we included a 60 second cadence measurement to better observe the dependence of cadence on occultation detection probability.

\begin{figure}
    \centering
    \includegraphics[width=0.5\textwidth]{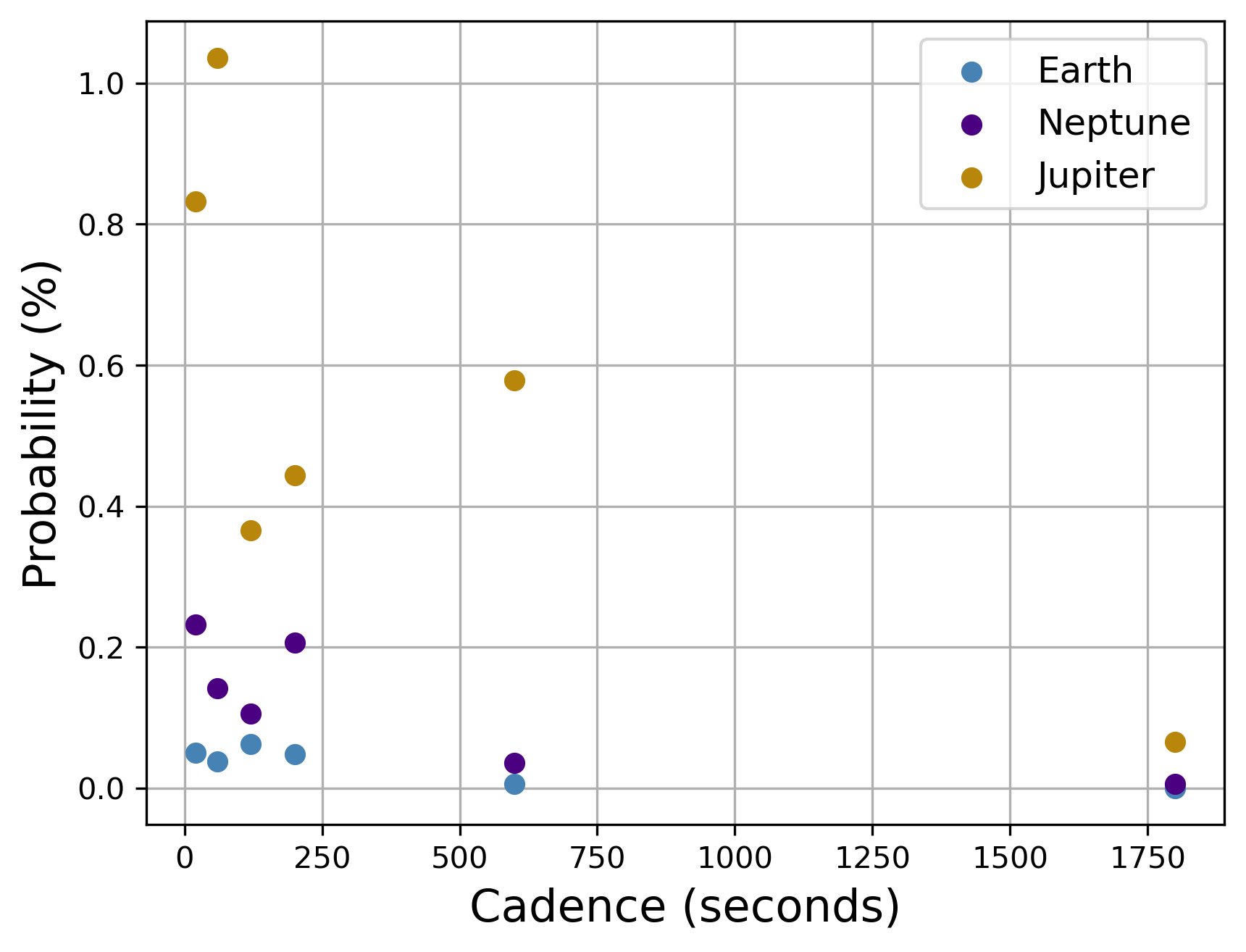}
    \caption{Single transit flare occultation probability as a function of the cadence of measurement. The cadence options are based on the TESS postage stamp (short cadence) and full-frame image (long cadence) data, which have changed throughout the mission lifetime: 20s, 60s, 120s, 200s, 600s, and 1800s. Again we use base parameters of $R_\star$ = 0.25R$_\odot$, a flare rate of 3 flares per day, $R_{\rm p} = 1R_\oplus, 1R_{\rm Neptune}, 1R_{\rm Jupiter}$, and a period of 7 days. For long cadence data, the probability becomes effectively zero because the occultation event is a shorter timeframe event. We require at least two data points during occultation to consider the event detectable.}
    \label{fig:cadence}
\end{figure}

The results of these simulations are shown in Figure~\ref{fig:cadence} where we see that shorter cadence data has an increased likelihood of detecting a flare occultation event. However, between the 60-second and 120-second cadence data, we do not see much change in their ability to detect occultation events. The overall trend in the probability is likely due to the duration of the occultations themselves. Occultations that are identified by our simulations as distinguishable are defined to be shorter than the overall flare length and occurring across multiple data points. The cadence then becomes important as not all occultation events could be distinguished at all cadences. While some flares may last upwards of hours, some may be on the order of a few minutes \citep{2022ApJ...926..204H} to seconds \citep{2022PASJ...74.1069A}. Flares of this timescale might be missed at certain cadences, let alone the occultations. Thus as the cadence of observation increases, it becomes less likely to detect as there are fewer photometry data points collected during the occultation, or it is possible to miss the occultation event altogether.  Additionally, if an occultation were to occur on longer cadence data and be detected, it would become increasingly difficult to identify. These events already have a tricky morphology that is obscured with higher cadences. Therefore, it is recommended when conducting occultation observations that the shortest cadence possible should be used. 

\subsection{Occultation Probabilities for Notable Systems}\label{subsec:notable_systems}

We demonstrate the probability of flare occultations for a number of real-world systems. For this sample, we have selected two well-known M dwarf exoplanet systems, TRAPPIST-1  and AU Mic, two double M dwarf eclipsing binaries - CM Draconis and GJ 3236, and identified systems of notable interest in the TOI catalog developed in \citet{2022MNRAS.512L..60H} . In order to calculate the probability of witnessing an occultation, we take into consideration the radius of the host/primary star, the radius of the secondary star or the transiting planet, the transit duration, the average flare rate, and the impact parameter for the transiting planet. For the systems gathered from \citet{2022MNRAS.512L..60H}, the impact parameter for many of the planets has yet to be clearly defined, so we assumed b = 0 for our probability predictions. However, it should be noted that in Fig~\ref{fig:probability_b} higher impact parameters yield greater occultation probabilities.

We calculate the occultation probability per transit and also per TESS sector, where we assume a TESS sector has 25.5 days. Finally, we calculate the occultation probability for all TESS data available up to and including cycle 5. CM Draconis is near the TESS continuous viewing zone, and hence has a substantial amount of data (16 sectors total). TRAPPIST-1, on the other hand, is near the ecliptic. It is yet to be observed but is scheduled for Cycle 6 (starting September 2023). Even without TESS data, our analysis is applicable to any photometric campaign with continuous monitoring and a cadence better than a few minutes.

\begin{table*}{}
    \centering
    \begin{tabular}{|c|c|c|c|c|c|c|c|c|c|}
    \hline
    System & $R_{\rm 1}$ & $R_{\rm 2}$ & b & FR & P & Prob (\%) & Prob (\%) & Prob (\%) & Num Sectors\\
    \hline
    & ($R_\odot$) & ($R_\odot$ or $R_\oplus$) & & (count/day) & (days) & (per transit) & (per sector) & (all TESS) &\\
    \hline
    \hline
        
    AU Mic b & 0.75  & 4.19 $R_{\oplus}$ & 0.26 & 5.54 & 8.463 & 0.38 & 1.1 & 2.3 &2\\
    AU Mic c & 0.75  & 2.79 $R_{\oplus}$ & 0.3 & 5.54 & 18.859 & 0.24 & 0.32 & 0.65 &2 \\
    TRAPPIST-1 b & 0.1192  & 1.116 $R_{\oplus}$ & 0.095 & 0.38 & 1.511 & 0.03  & 0.50 & N/A &0\\
    TRAPPIST-1 c & 0.1192  & 1.097 $R_{\oplus}$ & 0.109 & 0.38 & 2.422 & 0.046  &0.48 & N/A &0\\
    TRAPPIST-1 d & 0.1192  & 0.788 $R_{\oplus}$ & 0.063 & 0.38 & 4.049 & 0.028  & 0.18  & N/A &0\\
    TRAPPIST-1 e & 0.1192  & 0.92 $R_{\oplus}$ & 0.191 & 0.38 & 6.101 & 0.02  & 0.08 & N/A &0\\
    TRAPPIST-1 f & 0.1192  & 1.045 $R_{\oplus}$ & 0.312 & 0.38 & 9.207 & 0.026  & 0.07 & N/A &0\\
    TRAPPIST1 g & 0.1192  & 1.129 $R_{\oplus}$ & 	0.379 & 0.38 & 12.352 & 0.044 & 0.09 & N/A &0 \\
    TRAPPIST-1 h & 0.1192  & 0.755 $R_{\oplus}$ & 0.378 & 0.38 & 18.773 & 0.024 & 0.03 & N/A &0 \\
    TOI 176.01 & 0.32  & 10.432 $R_{\oplus}$ & 0 & 0.179 & 16.155 & 0.064 & 0.10 & 1.9 & 19 \\ 

    TOI 263.01 & 0.44  & 9.695 $R_{\oplus}$ & 0 & 0.025 & 0.557 & 0.01 & 0.45 & 0.71 & 2
    \\

    TOI 278.01 & 0.3  & 2.529 $R_{\oplus}$ & 0 & 0.076 & 0.299 & 0.006 & 0.51 & 0.79 & 2
    \\
    
    TOI 549.01 & 0.31  & 2.444 $R_{\oplus}$ & 0 & 0.227 & 0.516 & 0.02 & 0.98 & 2.5 & 3
    \\

    TOI 1779.01 & 0.31  & 10.978 $R_{\oplus}$ & 0 & 0.291 & 1.882 & 0.10 & 1.4 & 1.4 & 1
    \\

    CM Dra & 0.2534  & 0.2310 $R_{\odot}$ & 0.11 & 0.51 & 1.27 & 0.98 & 33 & 99.8 &15 \\
    GJ 3236 & 0.3729  &  0.3167 $R_{\odot}$ & 0.9  & 1.44  & 0.77  & 0.78 & 40.39& 87.37  &4 \\
    
    \hline
    \end{tabular}
    \caption{Occultation probabilities for a number of notable systems. They were derived after 50,000 transit simulations of each system using the stellar and planetary parameters in the above table. Each simulation operated on the assumption of a uniform flare distribution. The probabilities are calculated per transit, per TESS sector, and per total TESS lightcurve. Note that TRAPPIST-1 is yet to be observed by TESS.}
    \label{tab:known_systems_probabilities}
\end{table*}
As we can see in Table ~\ref{tab:known_systems_probabilities} for each of these systems, the probability of observing an occultation event in a single transit is quite low, even being a fraction of a percent at times. However, it is important to note this is the probability of an event occurring in a single transit, so with more data and transits observed, it will quickly rise. 
This can be seen in the later columns of Table ~\ref{tab:known_systems_probabilities} where the probabilities jump dramatically for the shorter period objects and those with ample observations already taken. These probabilities are the most important as 
they allow us to identify useful high-probability systems while also giving us a baseline to check against once observations have been taken and analyzed. By comparing the expected probabilities to the frequency of occultation events, it can provide insight into potential flaring distributions that may be occurring.

\section{A Search for Occulted Flares}\label{sec:methodology}

We implement our technique with TESS photometry. The methods here are generally applicable to short cadence photometric campaigns, but our specific work is with TESS. Given the results of Sect.~\ref{subsec:notable_systems} showing that eclipsing binaries have a significantly higher chance of flare occultations than transiting exoplanets, we apply our search techniques to the GJ 3236 and CM Draconis double M dwarf eclipsing binaries.

\subsection{Light Curve Processing}
To analyze the light curves we will be using data from the TESS survey. This data will be imported using the python package \emph{lightkurve} \citep{lightkurve}. We will select the shortest cadence data available because of the increased probability of observation that comes from short cadence data. As this is raw data we will also need to detrend the photometry, removing any stellar variation or noise from the telescope within the observations, before doing our fits. Any detrending in our analysis is done using the \textsc{Wotan} python package \citep{Hippke:2014} following the steps described in \citet{Martin2023CMDraconis}. 

\subsection{Exoplanet Fit}
Following the detrending of the light curves, the transits need to be fit. This is accomplished using the \textit{exoplanet} fitting program \citep{foreman-mackey2020}. The fits are performed using both the photometry data as well as the radial velocity data. Both data sets are fit simultaneously to ensure high accuracy in the photometry fits. For our purposes, the photometry curves are the more important fits, as the models will then be used to process the light curve further in order to isolate any occulted flares. For a more in depth look at this methodology look to \citet{Martin2023CMDraconis}.

\subsection{Identifying Flare Occultations}
Following the fitting of the photometry, the fits must be subtracted from the light curve.  The resulting light curve should be flat (i.e. devoid of out-of-eclipse variations like spot modulation and ellipsoidal variation) except for flares. This allows us to focus on only those that occurred during the transits. 

To identify targets we simply use visual inspection due to the small number of flares that occur during transits.
 This is reasonable since in this paper we are not trying to calculate any occultation statistics, but simply to look for any examples. Looking by eye lets us pick up the unique occultation signatures that might get missed by an automated algorithm. Since we are only looking for in-transit flares, we only have to cover $\sim 5-10\%$ of the light curve.

\subsection{No Candidates in GJ 3236}\label{subsec:GJ3236}

\begin{figure*}
    \centering
    \includegraphics[width =0.99\textwidth]{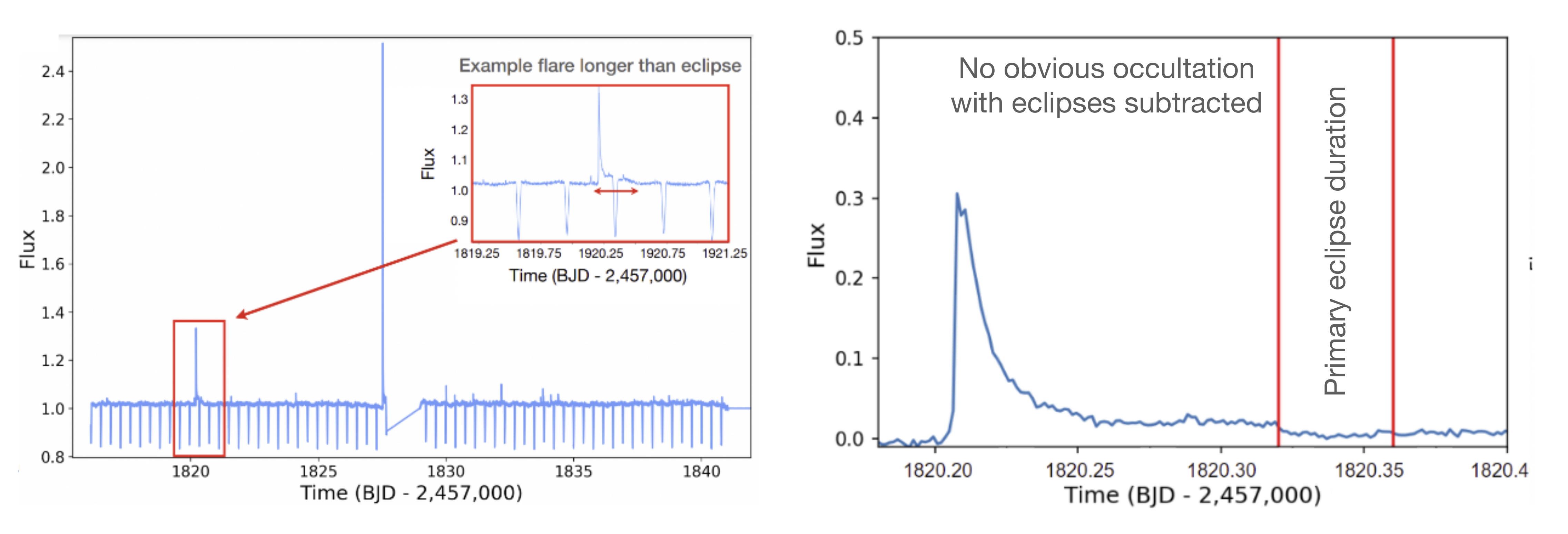}
    \caption{The photometry data collected for GJ 3236 in Sector 19 of TESS, which was used in our search for flare occultations. Many flares occurred near or during eclipse, with a large amplitude flare being highlighted that lasted for the entire duration of the eclipse. On the left shows the raw data. On the right shows the data with the eclipses removed, with no obviously visible flare occultation.}
    \label{fig:GJ3236_lightcurve}
\end{figure*}

GJ 3236 was observed by TESS in sector 19 and 25. We plot the light curve in Fig.~\ref{fig:GJ3236_lightcurve}. We noticed many flares had occurred during or just before eclipse, creating the possible conditions for an occultation event. There was also one example flare that was so long it lasted roughly four times the eclipse duration and covered an eclipse. However, after subtracting the eclipse we did not detect any morphology matching an occultation like Fig.~\ref{fig:flare_examples}. We visually analyzed all flares near eclipses and we did not find evidence of an occultation event occurring within either the sector 19 or sector 25 data. 

\subsection{One Candidate in CM Draconis}\label{subsec:CMDraconis}

\begin{figure*}
    \includegraphics[width = 0.99\textwidth]{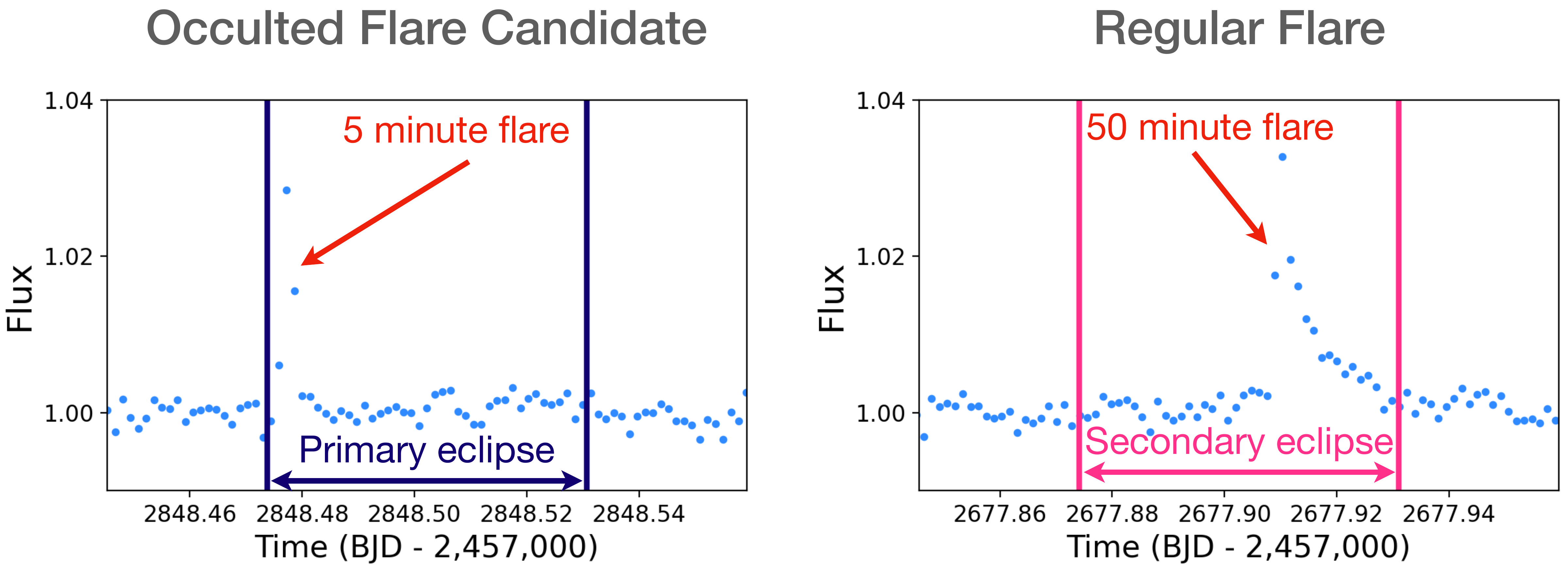}
    \caption{Two example flares on CM Draconis that occur during an eclipse. The flare on the left, occurring during a primary eclipse, lasts only five minutes. The flare on the right, occurring during a secondary eclipse, lasts $10\times$ longer despite having roughly the same amplitude. The very short decay time of the flare on the left may be due to an occultation event by the foreground star.}
    \label{fig:occulted_flare_comparison_example}
\end{figure*}

\begin{figure*}
    \centering
    \includegraphics[scale = 0.4]{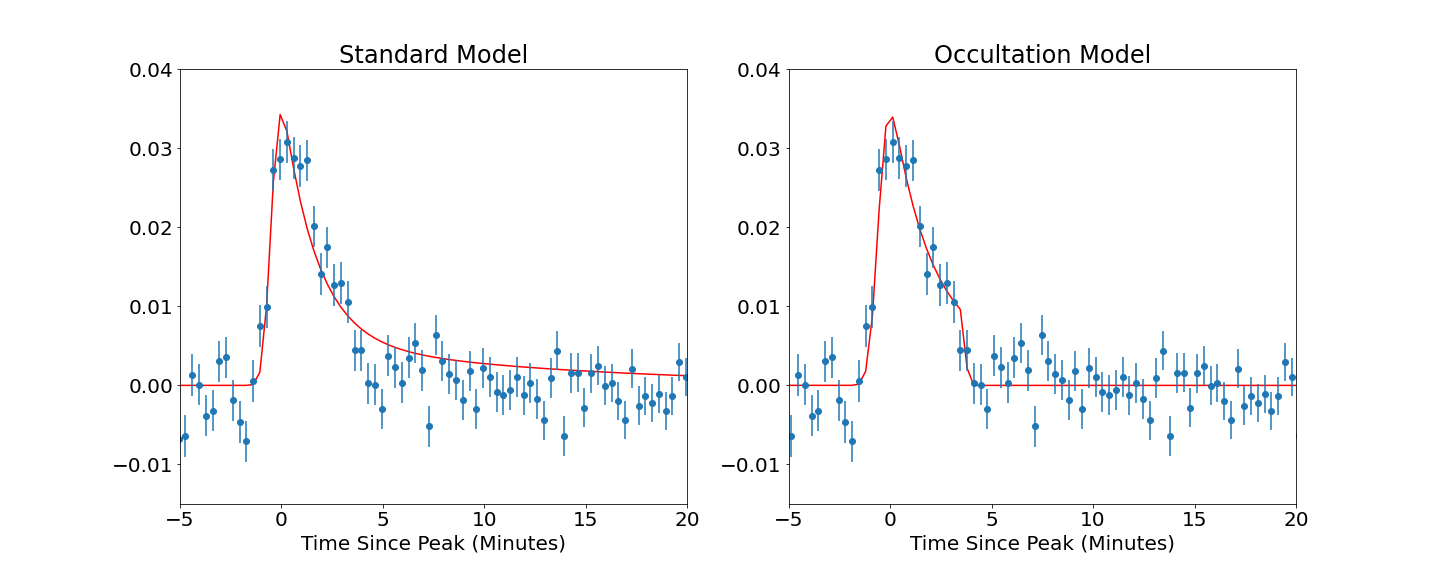}
    \caption{20-second cadence data for CM Draconis showing the occultation candidate pictured in Fig.~\ref{fig:occulted_flare_comparison_example} on the left. Here we see a greater definition of the sharp decline in flux that had occurred at around 4 minutes from the flare's peak. Shown in red is a model fit using the \citet{Mendoza2022} flare template, with the right panel displaying a flare morphology with an injected occultation. This candidate has a false positive chance of 4.8\% based on the comparisons of its FWHM and decay time to flares found within the \citet{2022ApJ...926..204H} sample.}
    \label{fig:20s_Occultation}
\end{figure*}

\begin{figure*}
    \includegraphics[width = 0.99\textwidth]{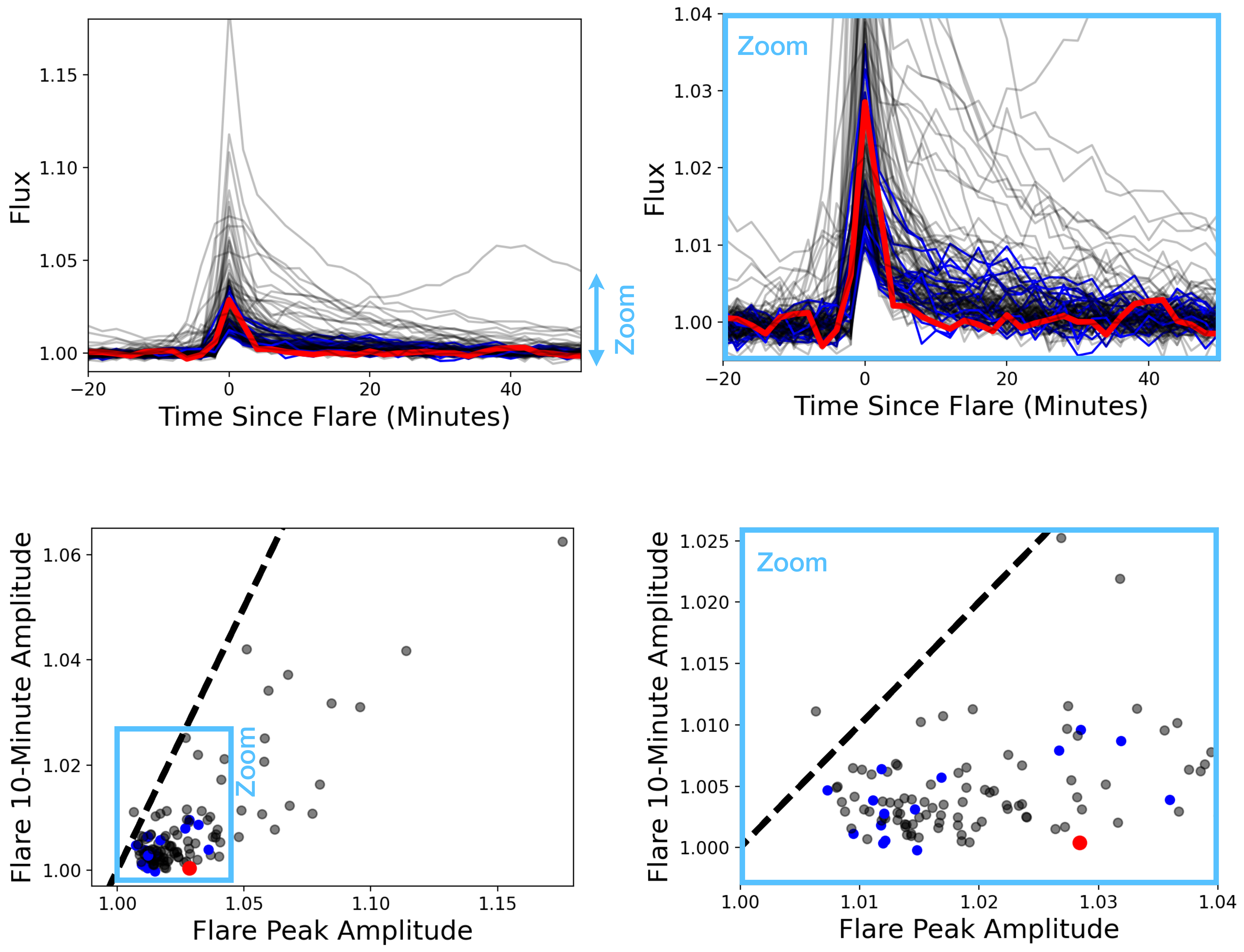}
    \caption{{\bf Top row:} 125 flares discovered in CM Draconis, stacked together and aligned to have a peak at $t=0$. The candidate occulted flare is highlighted in red. Flares occurring during the eclipse but likely not occulted are in blue. All other flares are in grey. The two top plots show the same information, but the top right is zoomed in on the occulted flare candidate, which is seen to drop in flux quicker than almost every other flare. {\bf Bottom row:} flare amplitude 10 minutes after peak as a function of the flare amplitude at its peak. The bottom right plot is zoomed on the blue square parameter space. Color-coding is the same as the top plots. Higher-amplitude flares tend to decay over a longer time, hence there is generally a positive correlation between the peak and 10-minute amplitude. However, for a given peak amplitude there is a significant spread in how long the decay takes. The candidate occulted flare definitely had a rapid decay, but it is not a large outlier.}
    \label{fig:occultation_candidate_analysis}
\end{figure*}

CM Draconis was studied in detail by \citet{Martin2023CMDraconis}. They discovered 125 flares in 12 TESS sectors. They proposed one possible flare occultation event, but did not study the event in detail. The candidate occultation is shown in  Fig.~\ref{fig:occulted_flare_comparison_example} on the left, highlighted in red, and in Figure~\ref{fig:20s_Occultation}. This flare occurs just as the primary eclipse begins and displays a unique morphology, in that it has a sharp rising phase, but rather than having a steady decay phase we see it immediately drop in flux. Overall the flare lasts only 5 minutes. For comparison, in Fig.~\ref{fig:occulted_flare_comparison_example} on the right we show another flare of a similar amplitude that occurred during an eclipse. This other flare was seemingly not occulted, since it displays a typical flare profile lasting  50 minutes.
The sharp rise and equally sudden decline of the candidate occulted flare matches morphology (b) from Fig.~\ref{fig:flare_examples}. In order to confirm a flare occultation event we must explore the significance of the deviation from standard flare morphology that the event displays. Looking at Fig.~\ref{fig:occultation_candidate_analysis}, which compares the occultation candidate's shape to each of the other flares that were analyzed. We see that this flare does decay much faster than a majority of the sample, but there are those that have similar decay rates, albeit slightly slower than the candidate. In addition, we look at the 10-minute amplitude of each of the flares, defined as the flux seen 10 minutes after the peak time of the flare.
When considering the 10-minute amplitude of these flares, we see a very wide distribution of flux after 10 minutes of decay.  The distribution is large, and the occultation candidate is an outlier on the lower end of the distribution. However, it is not significant enough to fully confirm its status as an occultation, leaving it as a candidate for now. We investigate newer 20-second cadence data of the candidate to explore if the flare is consistent with the signal expected from an occultation.

Looking at Fig.~\ref{fig:20s_Occultation} we see the occultation candidate, where at about 4 minutes following the peak time of the flare we see a quick decline in the flux during the decay phase of the flare. This drop in flux is where we believe a possible occultation may have occurred, and have therefore labeled it as a candidate for further analysis. In comparison to our predicted models this morphology is similar to that seen in Fig.~\ref{fig:flare_examples}B. We fit two models to the data shown in red in Fig.~\ref{fig:20s_Occultation} . On the left side is a standard flare model developed in \citet{Mendoza2022}, where the right side of the plot uses the same base model but with an occultation injected into the light curve. Performing a $\chi^2$ test we find that for the no occultation model the reduced $\chi^2 = 2.195$ where as the occultation model yields $\chi^2 = 1.508$. We then calculated the BIC for both the occultation model and the standard flare model with the occultation scoring -722.3 and the standard model having a BIC of -668.33. Taking both the reduced $\chi^2$ and BIC into account the model with the injected occultation does outperform the standard flare model. However the reduced $\chi^2$ is still greater than 1 for the occultation model, and therefore there is still some uncertainty as to whether this is truly an occultation. The lower chi-squared value compared to the flare model combined with the 99.81\% chance of observing an occultation strengthens the case for considering this event as a true flare occultation.

However, we still proceed cautiously and require further analysis to be done utilizing more complex modeling to say that an occultation truly occurred. One way to do this would be to use the fast flare template detailed in \citet{2022PASJ...74.1069A}. Additionally, it could be worth exploring a range of differing flare model parameters and baselines. In the case of the demonstration in Fig.~\ref{fig:20s_Occultation}, there is some variability in the baseline flux which may indicate the flare started at a lower flux than estimated. For our purposes, we found our chosen baseline appropriate, but in the case of a more detailed analysis, it may be required to test multiple baseline values.

\begin{figure}
    \centering
    \includegraphics[width = 0.4\textwidth]{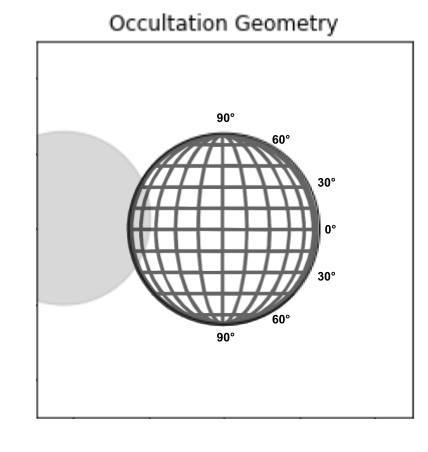}
    \caption{A depiction of the eclipse geometry for CM Draconis at the time of the candidate flare occultation in Figs.~\ref{fig:occulted_flare_comparison_example} and \ref{fig:20s_Occultation}. In gray is the secondary star, which is in the foreground as the flare occurred during a primary eclipse. The primary star, in the background, is shown with latitudes labeled. Due to the timing of the flare it would be constrained by the geometry above. The latitude range of the occultation would be between $\approx 30^{\circ}$ from the equator, assuming spin-orbit alignment for the tight eclipsing binary (assumption discussed in Sect.~\ref{subsec:caveats}).}
    \label{fig:geometry}
\end{figure}

While not confirmed as an occultation, we can use this candidate to show some analysis that could be done if it is indeed a flare occultation. The simplest conclusion that can be drawn is that this flare had occurred during a primary eclipse and, therefore, would have happened on the primary star. However, from the timing of the flare, we are able to predict where it may have occurred on this primary star as well. We simulated the eclipse and, in doing so, used the timing of the occultation to locate where the secondary star would be relative to the primary star at this point of the eclipse. We see in Fig.~\ref{fig:geometry} that only a small fraction of the primary star had been eclipsed at this point. With such a high radius ratio, had this occultation occurred later on, it would be impossible to constrain, but we see that the secondary star is only covering a range of latitudes $\pm$ 30 degrees from the equator at this point. This holds significance as in \citep{Ilin2021}, flares on M dwarfs had been constrained between 55 degrees and 81 degrees, pushing their locations closer to the poles. Whereas in our occultation candidate, we see evidence for an equatorial flare which would raise concerns for the habitability of M-dwarf planets if they tended to flare in this region as well.

Upon analyzing the rest of the photometry, this remained the only candidate for an occultation event, showcasing their rarity. Even so, this was sufficient for displaying how these types of events can be identified, where future follow-up surveys may detect more events in order to begin constraining the latitudes at which the flares are occurring. 

\section{Expectations for TESS}

We estimate the number of detectable occultation events within the existing TESS data archive. To estimate our sample we consult the TESS Eclipsing Binary Catalog\citep{2022yCat..22580016P} and the TOI flare catalog \citep{2022MNRAS.512L..60H}. Across each of these catalogs, there are a total of 4584 eclipsing binaries and 2250 TOIs respectively. However, for this work we only consider the M dwarf systems which cuts down the sample to 273 TOIs and 327 eclipsing binary systems.

We again trim this sample down as we are only concerned with flare stars, whose occurrence rates differ for the different types of M dwarfs. Thus we create two bins, of objects orbiting M0-M3 stars, and those orbiting M4-M6 stars. Following the flare star occurrence rates reported in \citet{2020AJ....159...60G}, and taking the median values for each bin, we can expect ~10\% of the M0-M3 and ~40\% of the M4-M6 to be active.  This gives us 24 TOIs and 26 eclipsing binaries in the M0-M3 bin and 14 TOIs and 29 eclipsing binaries in the M4-M6 bin.
\subsection{Assumptions}\label{sec:assumptions}
Using these systems we can then get an estimate for how many flare occultations we can expect across the TESS sample. In our estimations, we will be making several assumptions and simplifications to calculate the probability of witnessing an occultation. An in-depth analysis would require an in-depth exploration of each system's parameters which is beyond the scope of this work.

For the TOIs we will be considering 2 cases: they are entirely earth-like with radii of 1 R$_{\oplus}$ or they are all sub-Neptunes with radii of 4R$_{\oplus}$. We also assume that each system will have an impact parameter of b = 0. We then give a range of probabilities dictated by the period of the system with P = 1 day, and P = 10 days.

In the cases of the eclipsing binaries, we adopt the same period scheme, but our other assumptions differ slightly. For simplicity, we assume that the primary and secondary stars will have the same radii, and the eclipse will not be perfectly aligned. This means that rather than choosing an impact parameter of b = 0, we adopt a convention of b = 0.5, as grazing eclipses will be more useful as they do not occult all latitudes of the star.

Finally, for the stellar parameters, we adopt flare rates that were reported in Figure 11 of \citet{2020AJ....159...60G}. This gives a flare rate of 2 flares/day for the M0-M3 stars and 1 flare/day for the M4-M6 stars. Then for the stellar radii for the M0-M3 and M4-M6 bins, we adopt the radii reported in \citet{2013ApJS..208....9P} for an M2V and M5V star respectively. This gives the M0-M3 bin a primary star with radius R= 0.446 R$_{\odot}$ and the M4-M6 a star with radius R = 0.196 R$_{\odot}$.

\subsection{Probabilities and Expected Observations}

To calculate the probabilities of observing a single occultation event for each type of system within our bins we used the methodology detailed in section  
\ref{sec:probability} and the assumptions detailed in section \ref{sec:assumptions}. Additionally, we assume that at most each target will only have one sector of TESS observations, which were taken using a two-minute cadence. The results of these probability calculations can then be found in Table \ref{tesspredictions}

\begin{table*}{}
    \centering
    \begin{tabular}{|c|c|c|c|c|c|c|c|}
    \hline
    $R_{\rm 1}$ & $R_{\rm 2}$ & b & FR & P & Prob (\%) & Prob (\%) & Expected Observations\\
    \hline
    ($R_\odot$) & ($R_\odot$ or $R_\oplus$) & & (count/day) & (days) & (per transit) & (per sector) & Available TESS Data\\
    \hline
    \hline
     0.446 $R_{\odot}$  & 1 $R_{\oplus}$ & 0 & 2 & 1 & 0.034 & 0.86 & 0  \\
     0.446 $R_{\odot}$  & 1 $R_{\oplus}$ & 0 & 2 & 10 & 0.061  & 0.16 & 0\\
     0.446 $R_{\odot}$  & 4 $R_{\oplus}$ & 0 & 2 & 1 & 0.34 & 8.3 & 2\\ 
     0.446 $R_{\odot}$  & 4 $R_{\oplus}$ & 0 & 2 & 10 & 0.23 & 0.59 & 0\\
     0.446 $R_{\odot}$  & 0.446 $R_{\odot}$ & 0.5 & 2 & 1 & 1.4 & 51 & 13 \\
     0.446 $R_{\odot}$  & 0.446 $R_{\odot}$ & 0.5 & 2 & 10 & 1.3 & 6.5 & 2\\
     0.196 $R_{\odot}$  & 1 $R_{\oplus}$ & 0 & 2 & 1 & 0.017 & 0.43 &  0\\
     0.196 $R_{\odot}$  & 1 $R_{\oplus}$ & 0 & 2 & 10 & 0.074 & 0.19 & 0\\
     0.196 $R_{\odot}$  & 4 $R_{\oplus}$ & 0 & 2 & 1 & 0.27 & 6.7 & 1 \\
     0.196 $R_{\odot}$  & 4 $R_{\oplus}$ & 0 & 2 & 10 & 0.4 & 1.0 & 0 \\
     0.196 $R_{\odot}$  & 0.446 $R_{\odot}$ & 0.5 & 2 & 1 & 0.41 & 19 & 6\\
     0.196 $R_{\odot}$  & 0.446 $R_{\odot}$ & 0.5 & 2 & 10 & 0.93 & 4.7 & 1\\
     \hline
    
    \end{tabular}
    \label{tesspredictions}
    \caption{The above table lists the probabilities of witnessing an occultation for a number of different types of systems. Then, using the M0-M3 and M4-M6 bins of flare stars in the TESS sample, we report the expected number of occultations that would be observed, assuming all like systems had only one sector of data available.}
\end{table*}

We then used the probabilities of occultation detection to predict the number of observations we would expect for each scenario described at the beginning of this section. These predictions can also be seen in table \ref{tesspredictions}. We estimate that the majority of occultations will be found in short cadence observations of eclipsing binary systems, with 3-19 occultations present within current TESS observations. Up to 3 detectable occultation events by Neptune-sized planets are also likely within estimates. Our results show that occultations by Earthlike planets are very unlikely to have been captured by TESS, regardless of their orbital parameters, within a 28 day sector. However, these estimates are conservative as they do not take into consideration additional sector overlaps nor continued observations of TESS. 

\section{Discussion}\label{sec:discussion}
\subsection{Occultation Morphology}
In section 2 we set up a simulation to show how a flare occultation may look (Fig.~\ref{fig:flare_examples}). There were 3 distinct ways that the morphology could appear. Each of these models were built using simple flare models from \citet{Davenport2014}. In the case of X-ray flares used in the analysis of \citet{2007A&A...466..309S} these types of models would be sufficient as they do not display a large degree of temporal substructure. However in order to discuss the morphology of optical flare occultations, we must address that these flares display many complex behaviors \citep{2022ApJ...926..204H}. These complex behaviors can mimic occultation signals and can cause significant uncertainties when analyzing an occultation candidate. Moving forward, we address the different occultation morphologies we predict and the degeneracies that occur due to optical flares' complex behavior.

\subsubsection{Morphology A}
The morphology a) occurs when the flare is occulted immediately after the rising phase, but the occultation is short enough for some of the decay phase to be observed. This is usually associated with either smaller planets, long-lasting flares, or for short-period systems. The signal from such an event can serve as distinguishable evidence for an occultation but may be confused for sympathetic flaring. Sympathetic flaring occurs when two or more flares occur close together in time. As such, we may see that a short-duration flare occurs first, with a lower amplitude flare occurring afterward resulting in a light curve similar to that of morphology a). 
In addition to sympathetic flaring, with the release of TESS 20-second cadence data, \citet{2022ApJ...926..204H} found that a significant amount of flares had very complex substructures. Some of these substructures in particular could mimic the appearance of an occultation with morphology a).

The first of these substructures detailed in \citet{2022ApJ...926..204H} that could cause an issue is the quasi-periodic pulsations within M-dwarf optical flares. These pulsations can mimic an occultation as following a dip in the flux of a flare, the light curve intensifies and once again decays. These quasi-periodic pulsations occurred in a significant of the flare sample in \citet{2022ApJ...926..204H}, and typically had periods of 10 minutes or less, but some displayed even longer periods. There will be uncertainty in occultations of morphology a) due to these behaviors.

In analyzing flares of morphology a) it may be necessary to limit candidates to longer-lived flares that are completely occulted by the transiting body. This will create a single non-periodic drop in the flux received from the flare, where it is completely lost. As the signal returns, analysis of the remaining light curve should be done to ensure that no other significant pulsations occur within the light curve. It was also shown in \citet{2022ApJ...926..204H} that the pulsations often occur within the rising phase of the flare as well, which could be used to compare to the occultation signal. 
However, if the periodicity of the pulsations and occultation duration are on similar timescales it may become impossible to distinguish the events. Otherwise, an occultation candidate may still be found within flares displaying this behavior if its duration is significantly longer than other pulses observed in the flare. Difficulties in this situation may still be present as quasi-periodic pulsations can vary in both period and amplitude \citep{2022ApJ...926..204H} 

\citet{2022ApJ...926..204H} also made note of a "Peak-bump" flare. These flares still display a standard peak structure as shown in simple flare shapes, however, during the decay phase there is a secondary Gaussian bump that appears until it decays once more. Similar to the sympathetic flaring, this creates a single signal in the flare, which mimics the flare being occulted. To avoid confusion with 'Peak-bump' flares, candidates whose flux does not drop to 0 during the occultation will require extensive additional analysis to ensure that they are true occultations. An important consideration of the transit duration when modeling the geometry can assist in this effort, as the time difference of the peak to bump can be compared to the expected time the flaring region would be completely occulted. Another strategy to differentiate this behavior from an occultation of morphology a) would be to fit a standard \citet{Davenport2014} flare model with a secondary Gaussian fit as was done in \citet{2022ApJ...926..204H}. Then, perform a fit with an injected occultation and follow the procedures presented in this paper to gain additional confidence in the flare signal.

\subsubsection{Morphology b}
We also produced morphology b) as another type of occultation that may occur during the decay phase of the flare. This, again, can result in an identifiable signal, which is less likely to be mimicked by the complex flares identified in \citet{2022ApJ...926..204H}. With flares of morphology b) the signal will be a single sharp decline in flux shortly into the decay phase of the flare, which deviates greatly from the quasi-periodic pulsations of complex flares. However, general flare behaviors may still mimic these signals, as some flares display rapid decay, causing a sharp drop in flux following the peak of the flare. Rapidly decaying behavior could then create possible false positive occultation detections. Therefore, when considering a candidate that was completely occulted during the decay phase, it is vital to consider the other flares that had occurred in those observations to identify the significance of the deviation from standard flare morphology. 

\subsubsection{Morphology C}

Our simulations also produced morphology c) which is the result of a flare being occulted during its rising phase. The light curve produced here likely would not yield a detectable signal due to their resemblance to low-amplitude flares. In addition, in nearly half of the flares sampled in \citet{2022ApJ...926..204H} complex substructures were observed in the rising phase of the flare that could mimic these types of signals. Therefore, we do not recommend the usage of events with morphology c) for occultation candidates. Priority should be left to events with morphology b). Events with morphology a) can still be considered but will require significant analysis due to their similarities to complex flaring substructures. \\

\subsection{Occultation Probabilities}
Following the establishment of the flare morphology, we saw in our exploration of the probabilities that the detection of occultation events is quite rare, but this probability quickly rises with the more transits or eclipses that are observed. In addition, a few factors increase this probability as well, the first of which is the inherent flare rate of the host star. As this flare rate increases, we see a near-linear increase in the probability. A similar behavior is seen with the radius ratio, as the higher the radius ratio, the higher the probability of observing flare occultation. An interesting effect of the increased radius ratio is that it also increases the probability of observing an occultation event with higher impact parameters. This relationship does have a limit; however, as the transit moves to higher latitudes, the planet will begin to graze the star, which will decrease the probability of an observation. There is a trade-off with higher radius ratios, however: the larger the transiting body, the lower the precision with which we can localize a given flare.

This is due to the wider range of latitudes that are blocked out by larger planets, so in exchange for a larger chance of occulting a flare, we also see a decline in precision. 
We also saw with an increased period, there is a slight increase in the probability of witnessing an occultation event in a single transit. However, in the case of taking observations, which occurs across many days and hence may witness several transits, short-period planets would prove to be ideal. This is due to the many transits observed in the case of short periods, which quickly increase the odds of an identifiable event being detected. In longer periods it would be impractical to observe many transits and the overall probability of observing an occultation event would be lower than if many short-period transits were observed.\\

\subsection{Real Systems}
In this paper, we have simulated multiple stellar systems alongside a direct observation of two eclipsing binary systems. In observing the first eclipsing binary system, we looked at 
a single TESS sector and no evidence of an occultation event. However, in analyzing CM Draconis, we found a candidate for a flare occultation. The candidate was identified by its morphology resembling that produced by our simulations shown in Fig.~\ref{fig:flare_examples}c. In our analysis of the flare, we compared it to other flares of similar size seen in the CM Draconis system; however, it did not deviate greatly enough to be sure of an occultation.

If we consider the flares of similar amplitude in Table II of \citet{2022ApJ...926..204H}, we can compare to a larger sample of flares with similar magnitude to our occultation candidate. We selected flares first by those with TESS amplitudes of 0.03 or greater. Then the ratio of the flare decay time and FWHM time and FWHM duration were used as comparison metrics. For our occultation fit we estimate a FWHM duration of about 4 minutes, and the ratio of decay time to FWHM time as about 1.75. Comparing to the sample in \citet{2022ApJ...926..204H}, 4.8\% of flares had a decay time over FWHM time $\leq$ 1.75. However about 23\% of flares had FWHM durations of 4 minutes or less. This does favor the occultation candidate as being a true occultation, there is not a statistically significant difference to confirm it as an occultation.
In the future, as more 20-second cadence flares are detected this candidate could be revisited for further analysis.

In regards to our simulations, we saw that the probabilities we would predict are quite rare. Only the eclipsing binaries were able to exceed a single percent. However, as discussed previously, it would be harder to localize the flares within these systems. If we look at the TRAPPIST-1 system, we saw occultation probabilities at fractions of a percent. Yet, there are many opportunities for this system to produce an event. This is due to the 7 transiting planets around the star, along with each of them having short periods. In a typical TESS sector, the observing window is 27 days, and there are 3 planets that have periods of under 7 days. That would allow for at least 4 transits to be observed from each planet, with the maximum number being 18 transits within a single sector. Along with this many of the planets have similar impact parameters, meaning we have many checks to a similar range of latitudes. This would allow us to have an increased confidence in the flare distributions on this star if we were to perform an analysis of it. This shows us that even with small probabilities for observation for a given system, if there are other favorable parameters, this technique can still be a powerful tool.

\subsection{Expectations for Occultations}

Our simulation results show that multiple occultations have likely already been captured by TESS. In the case of the most conservative estimates, there is a possibility no planetary system contains an occultation event, but 3 would have been captured by eclipsing binaries. Our optimistic estimates place an upper limit on the number of occultations to be 22. To produce these estimates we also found that earth-like systems are unlikely to result in an occultation in a reasonable observation length. As such we recommend dedicated searches for flare occultations focus on systems with larger transiting planets to eclipsing binary systems. 

It is also important to address that while our estimates of occultations that exist within the TESS data set are low, some of our assumptions may cause us to underestimate the amount of available data. One such estimate is the number of TESS sectors that each system has been observed in. As we have demonstrated in section \ref{sec:probability}, the amount of existing data greatly increases the likelihood of detecting an observation. Multiple systems within our bins have likely been observed for multiple TESS sectors, which would raise the chances of an occultation. In addition, for each of the planetary systems we set the impact parameter to be zero. In the case of earth-sized planets, this would have little impact, but as the radius increases impact parameter plays more of a role. Specifically for the Neptune-sized planets, a higher impact parameter yields a larger probability of an occultation event occurring.

Our estimates also assume very basic system parameters to obtain an initial outlook on implementing this technique. There exists a multitude of systems with more favorable conditions for producing an occultation event such as CM Draconis and GJ 3236. Even if the number of occultations remains small any confirmed occultation will be extremely valuable for the characterization of M dwarf flaring latitudes. As such our understanding of these events is limited. Moving forward, when assessing the optical photometry of active datasets, it would still be important to check for these events, especially within those with short periods or large radius ratios. 

\subsection{Caveats}\label{subsec:caveats}

Before implementing this technique on a larger scale, we must address its caveats. The first of which to consider is the effect of spin-orbit alignment. This is when the exoplanet's orbital axis is aligned with the stellar spin axis. In this scenario, the exoplanet will be passing over a tightly constrained range of latitudes. For our analysis, we have been operating under the assumption that the exoplanets and stars are spin-orbit aligned, which is the case for most exoplanet systems. However, there do exist systems that are significantly misaligned, which would make constraining the latitude of the flare extremely difficult. As an example, the exoplanet may transit diagonally across the star, cutting across latitudes from 0 to 90 degrees. With such a range it would be nearly impossible to easily constrain the latitude the flare had occurred at. In order to combat this, extremely precise transit geometry modeling would be required to determine the latitudes covered at the exact moment of occultation. Thus, we recommend to use targets that are well-aligned with their stellar hosts when searching for occultation events.

In Section 7.1, we covered some of the degeneracies that may arise from differing flare morphology compared to occultation events. 
However, these are not the only things that may cause confusion. For occultation type (a), one of the major sources of uncertainty is sympathetic flaring. This occurs with two flares whose peaks occur close to each other in time. Efforts can be made to calculate the probability of sympathetic flaring events. Occultations with sharp signals will display features that tend to favor occultations over sympathetic flaring. Additionally, careful analysis of the light curve must be done as well, to ensure there was not a false signal was not produced. This difficulty may arise when considering the nature of the eclipse modeling which fits to an average of all of the eclipses available in the photometry data. However, this effect should not be considered to the same weight as sympathetic flaring. 

False signals may also occur for morphology (b) which has an absence of a decay phase, and resembles a spike in the light curve. Such signals may also appear in photometry as the result of events such as an asteroid crossing in view of the telescope, reflecting additional light creating a spike in the photometry. A similar effect can also be the result of a cosmic ray being detected as well. This caveat can be remedied with short cadence observations as with the cosmic ray the detection would result in one anomalous data point, but with many data points in the rising phase of the flare one would be able to rule out additional sources producing the spike in the photometry. 

Our criterion is dependent on unusual morphology, especially in the case of a flare being occulted at the end of its decay phase. Thus, we must consider the chances of unusually short standard flares. To counter this we must take into consideration other flares of similar amplitude to those identified as occultation candidates. Looking to Fig.~\ref{fig:occultation_candidate_analysis}, we compared the peak amplitude, and that seen after 10 minutes from the peak amplitude. In this case we are also cautious with confirming the occultation candidate as a true observation as the 10-minute amplitude is not a great outlier from similar amplitude flares. Statistical tests should be done regarding the duration of any occultations to that of other flares to show significant differences between an occultation candidate and short-lived flares. When considering shorter flares, we may also utilize the plateaued peak structure detailed in \citet{2022PASJ...74.1069A}. Rather than having a sharp peak in the flux, it levels out around the peak, which is an effect that can be seen in Fig \ref{fig:20s_Occultation}. These flares do still display a standard exponential decay, however, so occultations can be considered in those with durations one the order of a few minutes and greater, as with the flare in this analysis.

One final caveat is that the best candidates for flare occultations are eclipsing binaries. This then poses the question of if flare latitudes in eclipsing binaries are representative of flare latitudes in single stars. \citet{Simon1980,vandenOord1988,Gunn1997} proposed that in a tight binary magnetic field lines may traverse between the two stars, impacting stellar activity in a way that does not exist in single stars. Sethi et al. (in prep) show that tight eclipsing binaries have active longitudes of starspots near the sub- and anti-stellar points. Whether or not there are active longitudes and/or latitudes of flares in tight binaries, remains an open question.

\subsection{Transiting Planets vs Eclipsing Binaries}\label{subsec:EBs_vs_planets}

Discussed briefly in Section~\ref{sec:parameter_effects} was the different advantages that eclipsing binaries and exoplanets have when using this technique. Expanding upon that, eclipsing binaries are extremely useful in witnessing occultations as they cover a great area of their primary star leading to greater occultation probability. However, the greater occultation probability comes at a loss of precision. In the example shown in Section~\ref{sec:methodology}, we saw a case where we could heavily constrain an occultation event due to it occurring so early on in the eclipse. In this case, we were fortunate, and would be possible in other eclipsing binary systems to have this timing, but it is not reliable to happen consistently.
Eclipsing binaries do have another advantage as they are more likely to be tidally spin-orbit aligned than exoplanets, albeit based on minimal preliminary evidence \citep{triaud2017,Hodzic2020}. As such it would give a greater opportunity for constraining the latitudes which are being covered, alleviating one of the caveats to this technique discussed earlier.

\section{Conclusion}\label{sec:conclusion}
In this paper, we detail the applicability of flare occultation events as a method to identify flare latitudes in optical photometry.
Through the identification of multiple of these events, we can begin to work towards a general understanding 
of the preferential flaring regions on M dwarfs, which is a problem that must be addressed if we are to properly gauge the habitability of planets in orbit around them.

Through our simulations, we have predicted different flare morphologies that may be produced by these events to help guide future observations and research done using this technique. A caveat to this is that some morphologies may prove to be difficult to confirm as an occultation, yet the most common types are valuable assets to identify these events. In addition to this, we have predicted the probability of observing an occultation event given various parameters and for real systems. These events are rare, but identifiable, especially as more data comes in for these systems. Within the current dataset, we place a limit of 3-22 occultations that may already be identifiable.

We used this technique on two systems to search for these events and identified an occultation candidate within the CM Draconis system. By comparing it to other flares that occurred within the same observation window, we were then able to gauge its significance through two flare models but were not able to confirm it as an occultation. Despite this, we were able to detail the steps that can be taken to constrain an occultation event. Using its timing shortly into the eclipse, we were then able to provide a constraint on which star it occurred on, and the probable region in where it would have had to occured. This candidate, which if confirmed, would provide an additional data point in the small number of constrained flares on M dwarfs which have thus far been shown to occur on higher latitudes.

In searching for occulted flares, habitable zone planets are of high interest, as are their atmospheres. Thus any life would be susceptible to the effects of increased flare rates. 
For M dwarfs in particular, planets within the habitable zone display a variety of periods ranging from  17-148 days, depending on the host star. This value was derived using Kepler's third law and the snow line distances described in \citet{Childs_2022}. However, rocky planets are unlikely to produce occultation events. Luckily, a majority of TESS planets have radii >4.0 R$_{\oplus}$ \cite{2018ApJS..239....2B}. This is beneficial as larger systems can increase the odds of witnessing these events. As such, many systems within the TESS catalog are ready to be checked for flare occultations. Using the methodology described within this paper, this technique can serve as an additional tool in the mission to constrain the flaring regions on M dwarf stars.

\section*{Data Availability Statement}

All simulation and observational data may be provided upon reasonable request to the authors.

\section*{Acknowledgements}
Support for this work was provided by NASA through the NASA Hubble Fellowship grant HF2-51464 awarded by the Space Telescope Science Institute, which is operated by the Association of Universities for Research in Astronomy, Inc., for NASA, under contract NAS5-26555. This research was carried out in part at the Jet Propulsion Laboratory, California Institute of Technology, under a contract with the National Aeronautics and Space Administration (80NM0018D0004). RRM acknowledges support from the Presidential Fellowship granted by The Ohio State University. We would also like to thank our anonymous referees for their helpful feedback in producing this work.

\bibliographystyle{mnras}
\bibliography{Occultation.bib} 


\bsp	
\label{lastpage}
\end{document}